\documentclass{aa}  
\usepackage{graphicx}

\usepackage{txfonts}
\usepackage[labelfont=bf,list=true]{subcaption}
\begin{document}

   \title{Pluto's ephemeris from ground-based stellar occultations (1988-2016)}

   \author{J. Desmars\inst{1}
          \and
          E. Meza\inst{1}
          \and
          B. Sicardy\inst{1}
         \and
          M. Assafin\inst{2}
          \and
         J.~I.~B. Camargo\inst{3}
          \and
         F. Braga-Ribas\inst{4,3,1}
         \and 
         G. Benedetti-Rossi\inst{3} 
          \and
         A. Dias-Oliveira\inst{5,3}
          \and
         B. Morgado\inst{3}
          \and
         A. R. Gomes-Júnior\inst{6,2} 
          \and
         R. Vieira-Martins\inst{3}
         \and
         R. Behrend\inst{7}
         \and
          J.~L. Ortiz\inst{8}
          \and
         R. Duffard\inst{8} 
          \and
         N. Morales\inst{8} 
          \and
         P. Santos Sanz\inst{8}
          }

   \institute{LESIA, Observatoire de Paris, Universit\'e PSL, CNRS, Sorbonne Universit\'e, Univ. Paris Diderot, Sorbonne Paris Cit\'e, 5 place Jules Janssen, 92195 Meudon, France\\
              \email{josselin.desmars@obspm.fr}
             \and 
             Observat\'orio do Valongo/UFRJ, Ladeira Pedro Antonio 43, Rio de Janeiro, RJ 20080-090, Brazil
             \and 
             Observat\'orio Nacional/MCTIC, Laborat\'orio Interinstitucional de e-Astronomia-LIneA and INCT do e-Universo,
             Rua General Jos\'e Cristino 77, Rio de Janeiro CEP 20921-400, Brazil
             \and 
             Federal University of Technology - Paran\'a (UTFPR/DAFIS), Rua Sete de Setembro 3165, 
             CEP 80230-901 Curitiba, Brazil
             \and 
             Escola SESC de Ensino Médio, Avenida Ayrton Senna, 5677, Rio de Janeiro - RJ, 22775-004, Brazil
             \and 
             UNESP - São Paulo State University, Grupo de Dinâmica Orbital e Planetologia, CEP 12516-410, Guaratinguetá, SP 12516-410, Brazil
             \and 
             Geneva Observatory, 1290 Sauverny, Switzerland
             \and 
             Instituto de Astrof\'{\i}sica de Andaluc\'{\i}a (IAA-CSIC). Glorieta de la Astronom\'{\i}a s/n. 18008-Granada, Spain     
            }

   \date{Received December 21, 2018; accepted March 08, 2019}

 
  \abstract
   {From 1988 to 2016, several stellar occultations have been observed to characterize Pluto's atmosphere and its evolution \citep{mez18}. From each stellar occultation, an accurate astrometric position of Pluto at the observation epoch is derived. These positions mainly depend on the position of the occulted star and the precision of the timing. }
   {We present Pluto's astrometric positions derived from 19 occultations from 1988 to 2016 (11 from \cite{mez18} and 8 from other publications). Using Gaia DR2 for the positions of the occulted stars, the accuracy of these positions is estimated to 2-10~milliarcsec depending on the observation circumstances. From these astrometric positions, we derive an updated ephemeris of Pluto's system barycentre using the NIMA code \citep{des15}.}
   {The astrometric positions are derived by fitting the occultation's light curves by a model of Pluto's atmosphere. The fits provide the observed position of the body's centre for a reference star position. Other publications usually provide circumstances of the occultation such as the coordinates of the stations, the timing, and the impact parameter (i.e. the closest distance between the station and the centre of the shadow). From these parameters, we use a procedure based on the Bessel method to derive an astrometric position. }
   {We derive accurate Pluto's astrometric positions from 1988 to 2016. These positions are used to refine the orbit of Pluto'system barycentre providing an ephemeris, accurate to the milliarcsec level, over the period 2000-2020, allowing better predictions for future stellar occultations.}
   {}


   \maketitle
%

\section{Introduction}
Stellar occultation is a unique technique to obtain the physical parameters of distant objects or to probe their atmosphere and surroundings. 
For instance, \cite{mez18} have used 11 stellar occultations by Pluto from 2002 to 2016 to study the evolution of Pluto's atmosphere.
Meanwhile, occultations allow an accurate determination of the relative position of the body's centre compared to the position of the occulted star, leading to an accurate astrometric position of Pluto at the time of occultation if the star position is also known accurately. 

The accuracy of the body's position mainly depends on the knowledge of the shape and the size of the body, the modelling of the atmosphere, the precision of the timing system, the velocity of the occultation, the exposure time of the camera, the precision of the  stellar position, and the magnitude of the occulted star. Since the publication of the Gaia catalogues in September 2016 for the first release \citep{gai16} and moreover with the second release in April 2018 \citep{gai18} including proper motions and trigonometric parallaxes of the stars, the precision of the stellar catalogues can now reach a tenth of a milliarcsec (mas). For comparison, before Gaia catalogues, the precision of stellar catalogues such as UCAC2 \citep{zac04} or UCAC4 \citep{zac13}, was about 50 to 100~mas per star with also zonal errors. 
With Gaia, the precision of positions deduced from occultations is expected to be around few mas, taking into account the systematic errors.  Thanks to the accuracy of the GaiaDR2 catalogue, occultations can provide the most accurate astrometric measurement of a body in the outer solar system. 
 
 In this paper, we present the astrometric positions we derived from occultations presented in \cite{mez18} (Sect.~\ref{Ss:astrometryMeza}) and in other publications (Sect.~\ref{Ss:astrometryOther}). 
We detail a method to derive astrometric positions from other publications, knowing the circumstances of occultations: timing and impact parameter (Appendix). 
Finally, we present in Section~\ref{S:NIMA} a refined ephemeris of Pluto's system barycentre and we discuss the improvement in the predictions of future occultations by Pluto at a mas level accuracy and as well as the geometry of past events (Sect.~\ref{S:discussion}).


\section{Astrometric positions from occultations}\label{S:astrometry}

\subsection{Astrometric positions from occultations in \cite{mez18}}\label{Ss:astrometryMeza}

\cite{mez18} provide 11 occultations by Pluto from 2002 to 2016. Beyond the parameters related to Pluto's  atmosphere, another product of the occultations is the astrometric position of the body. From the geometry of the event, we determine the position of the Pluto's centre of figure $(\alpha_c,\delta_c)$. This position corresponds to the observed position of the object at the time of the occultation for a given star position $(\alpha_s,\delta_s)$. In particular, the position of the body we derive, only depends on the star position. Before Gaia catalogues, we determined the star position with our own astrometry. Table~\ref{T:offsetEM} gives the position of Pluto's centre and its precision we derived from the geometry of the occultation and the corresponding star position from our astrometry. With Gaia, the astrometric position of Pluto's centre can be refined by correcting the star position with the Gaia DR2 star position with the relations:

\begin{align}
\alpha & = \alpha_c +  \alpha_{GDR2}- \alpha_s \\
\delta & = \delta_c + \delta_{GDR2} - \delta_s 
\end{align}

This refined position only depends on the Gaia DR2 position which is much more accurate than previous catalogues or our own astrometry. The associated position of the occulted stars in Gaia DR2 catalogue ($\alpha_{GDR2},\delta_{GDR2}$) are listed in Table~\ref{T:stars-gaia}. The positions take into account the proper motions and the parallax from Gaia DR2. The table also presents the Gaia source identifier and the estimated precision of the star position in right ascension and declination, at the time of the occultation, taking into account precision of the stellar position and the proper motions as given in GaiaDR2, for all the occultations studied in this paper.  

\begin{table*}[htp!]
\begin{center}
\caption{Date, timing, position of Pluto's centre deduced from the geometry and the precision, coordinates of the occulted star used to derive the astrometric positions of occultations by Pluto studied in \cite{mez18}.}
\begin{tabular}{crcrcrr}
\hline
Reference date  & \multicolumn{4}{c}{Pluto's centre position}  & \multicolumn{2}{c}{position of star}    \\
 & right ascension & $\sigma_\alpha$ & declination & $\sigma_\delta$  & right ascension & declination \\ 
 & \multicolumn{1}{c}{$\alpha_c$} & (mas) & \multicolumn{1}{c}{$\delta_c$} & (mas)  & \multicolumn{1}{c}{$\alpha_s$} & \multicolumn{1}{c}{$\delta_s$}  \\ 
\hline
2002-08-21 07:00:32 & 16h58m49.4393s & 0.2 & -12$^{\circ}$51'31.944" & 0.1 & 16h58m49.4360s & -12$^{\circ}$51'31.920"  \\
2007-06-14 01:27:00 & 17h50m20.7368s & 0.1 & -16$^{\circ}$22'42.210" & 0.2 & 17h50m20.7392s & -16$^{\circ}$22'42.210"  \\
2008-06-22 19:07:28 & 17h58m33.0303s & 0.2 & -17$^{\circ}$02'38.504" & 0.2 & 17h58m33.0138s & -17$^{\circ}$02'38.349"  \\
2008-06-24 10:37:00 & 17h58m22.3959s & 0.1 & -17$^{\circ}$02'49.177" & 0.7 & 17h58m22.3930s & -17$^{\circ}$02'49.349"  \\
2010-02-14 04:45:00 & 18h19m14.3681s & 0.2 & -18$^{\circ}$16'42.125" & 0.5 & 18h19m14.3851s & -18$^{\circ}$16'42.313"  \\
2010-06-04 15:34:00 & 18h18m47.9476s & 0.3 & -18$^{\circ}$12'51.922" & 1.3 & 18h18m47.9349s & -18$^{\circ}$12'51.794"  \\
2011-06-04 05:42:00 & 18h27m53.8235s & 0.3 & -18$^{\circ}$45'30.741" & 0.3 & 18h27m53.8249s & -18$^{\circ}$45'30.725"  \\
2012-07-18 04:13:00 & 18h32m14.6748s & 0.1 & -19$^{\circ}$24'19.307" & 0.1 & 18h32m14.6720s & -19$^{\circ}$24'19.295"  \\
2013-05-04 08:22:00 & 18h47m52.5333s & 0.1 & -19$^{\circ}$41'24.403" & 0.1 & 18h47m52.5322s & -19$^{\circ}$41'24.374"  \\
2015-06-29 16:02:00 & 19h00m49.7122s & 0.1 & -20$^{\circ}$41'40.399" & 0.1 & 19h00m49.4796s & -20$^{\circ}$41'40.778"  \\
2016-07-19 20:53:45 & 19h07m22.1164s & 0.1 & -21$^{\circ}$10'28.242" & 0.4 & 19h07m22.1242s & -21$^{\circ}$10'28.445"  \\
\hline
\end{tabular}
\label{T:offsetEM}
\end{center}
\end{table*}

\begin{table*}[htp!]
\begin{center}
\caption{Gaia DR2 source identifier, right ascension and declination and their standard deviation (in mas) at epoch and magnitude of the stars of the catalogue Gaia DR2 involved in occultations presented in this paper. }
\begin{tabular}{ccrrrrr}
\hline
Date  & Gaia source identifier & right ascension & declination & $\sigma_\alpha$ & $\sigma_\delta$ & Gmag \\
\hline
1988-06-09 & 3652000074629749376 & 14h52m09.962000s & +00$^{\circ}$45'03.30297" &       2.14 &       2.06 &       12.1\\
2002-07-20 & 4333071455580364160 & 17h00m18.029957s & -12$^{\circ}$41'42.01220" &       1.12 &       0.73 &       12.6\\
2002-08-21 & 4333042833914281856 & 16h58m49.431538s & -12$^{\circ}$51'31.85910" &       1.87 &       1.12 &       15.4\\
2006-06-12 & 4124931567980280832 & 17h41m12.074271s & -15$^{\circ}$41'34.47421" &       0.63 &       0.49 &       14.7\\
2007-03-18 & 4144912550502784384 & 17h55m05.699098s & -16$^{\circ}$28'34.36682" &       0.74 &       0.60 &       14.8\\
2007-06-14 & 4147858103406546048 & 17h50m20.744804s & -16$^{\circ}$22'42.22719" &       0.83 &       0.73 &       15.3\\
2008-06-22 & 4144621347334603520 & 17h58m33.013236s & -17$^{\circ}$02'38.39643" &       0.67 &       0.54 &       12.3\\
2008-06-24 & 4144621244254585728 & 17h58m22.390423s & -17$^{\circ}$02'49.36558" &       0.93 &       0.78 &       15.6\\
2010-02-14 & 4096385295578625536 & 18h19m14.378482s & -18$^{\circ}$16'42.35590" &       0.50 &       0.42 &       10.3\\
2010-06-04 & 4096389556186605568 & 18h18m47.930034s & -18$^{\circ}$12'51.82967" &       0.37 &       0.31 &       14.8\\
2011-06-04 & 4093175335706340480 & 18h27m53.819996s & -18$^{\circ}$45'30.78871" &       0.62 &       0.50 &       16.4\\
2011-06-23 & 4093163211131448704 & 18h25m55.479351s & -18$^{\circ}$48'07.09094" &       0.35 &       0.31 &       14.0\\
2012-07-18 & 4092849712861519360 & 18h32m14.673688s & -19$^{\circ}$24'19.34329" &       0.19 &       0.17 &       14.4\\
2013-05-04 & 4086200313156846336 & 18h47m52.531982s & -19$^{\circ}$41'24.39714" &       0.10 &       0.09 &       14.2\\
2014-07-23 & 4085914882468876672 & 18h49m31.736687s & -20$^{\circ}$22'23.82473" &       0.21 &       0.19 &       17.2\\
2014-07-24 & 4085914745029913216 & 18h49m26.511650s & -20$^{\circ}$22'36.98627" &       0.39 &       0.35 &       18.1\\
2015-06-29 & 4084956039611370112 & 19h00m49.474124s & -20$^{\circ}$41'40.81016" &       0.04 &       0.04 &       12.0\\
2016-07-19 & 4082062610353732096 & 19h07m22.117772s & -21$^{\circ}$10'28.43508" &       0.05 &       0.05 &       13.9\\
\hline
\end{tabular}
\tablefoot{The coordinates and their precision are provided for the epoch of the occultation taking into account the proper motions and the parallax, and their precision.}
\label{T:stars-gaia}
\end{center}
\end{table*}

Finally, Table~\ref{T:Pluto-positions} provides the absolute position in right ascension and declination of Pluto's centre derived from the geometry and from stellar positions of Gaia DR2. The residuals related to JPL ephemeris\footnote{DE436 is a planetary ephemerides from JPL providing the positions of the barycentre of the planets, including the barycentre of Pluto's system. It is based on DE430 \citep{DE436}. PLU055 is the JPL ephemeris providing the positions of Pluto and its satellites related to the Pluto's system barycentre, developed by R.Jacobson in 2015 and based on an updated ephemeris of \cite{bro15}: \url{https://naif.jpl.nasa.gov/pub/naif/generic_kernels/spk/satellites/plu055.cmt}} DE436/PLU055 are also indicated as well as the differential positions between Pluto and Pluto's system barycentre used to refine the orbit (see Sect.~\ref{S:NIMA}). A flag indicates if the position is used in the NIMAv8 ephemeris determination. Finally, the reconstructed paths of the occultations are presented in Fig.~\ref{F:paths}.

\begin{table*}[htp!]
\begin{center}
\caption{Right ascension and declination of Pluto deduced from occultations, residuals (O-C) in mas related to JPL DE436/PLU055 ephemeris, and differential coordinates (PLU-BAR) between Pluto and Pluto barycentre system position from PLU055 ephemeris.}
\begin{tabular}{cccrrrrcc}
\hline
  & \multicolumn{2}{c}{Pluto's coordinates} &  \multicolumn{2}{c}{O-C (mas)} &  \multicolumn{2}{c}{PLU-BAR (mas)} \\
date (UTC) & right ascension & declination & $\Delta\alpha\cos(\delta)$ & $\Delta\delta$ & $\Delta\alpha\cos(\delta)$ & $\Delta \delta$ & flag & reference\\ 
\hline
  1988-06-09 10:39:17.0 & 14h52m09.96347s & +00$^{\circ}$45'03.1506" &         19.9 &        -33.5 &       -8.8 &       79.6 & * & \citet{mil93} \\
  2002-07-20 01:43:39.8 & 17h00m18.03018s & -12$^{\circ}$41'41.9934" &          7.7 &         -4.4 &      -52.9 &       24.7 & * & \citet{sic03} \\
  2002-08-21 07:00:32.0 & 16h58m49.43477s & -12$^{\circ}$51'31.8833" &         20.6 &        -10.4 &      -51.2 &       48.8 & * & This paper \\
  2002-08-21 07:00:32.0 & 16h58m49.43442s & -12$^{\circ}$51'31.8820" &         15.4 &         -9.1 &      -51.2 &       48.8 & * & \citet{ell03} \\
  2006-06-12 16:25:05.7 & 17h41m12.07511s & -15$^{\circ}$41'34.5896" &          9.8 &         -0.4 &      -47.0 &      -40.8 & * & \citet{you08} \\
  2007-03-18 10:59:33.1 & 17h55m05.69430s & -16$^{\circ}$28'34.0886" &         10.7 &          0.8 &       67.1 &      -39.4 & * & \citet{per08} \\
  2007-06-14 01:27:00.0 & 17h50m20.74243s & -16$^{\circ}$22'42.2275" &         14.7 &         -1.8 &       -5.2 &       89.8 & * & This paper \\
  2008-06-22 19:07:28.0 & 17h58m33.02976s & -17$^{\circ}$02'38.5534" &         14.0 &          0.0 &      -59.3 &      -23.3 & * & This paper \\
  2008-06-24 10:37:00.0 & 17h58m22.39339s & -17$^{\circ}$02'49.1932" &         17.6 &          8.1 &      -35.4 &       89.6 & * & This paper \\
  2010-02-14 04:45:00.0 & 18h19m14.36152s & -18$^{\circ}$16'42.1678" &         15.2 &          3.1 &      -65.4 &       55.6 & * & This paper \\
  2010-06-04 15:34:00.0 & 18h18m47.94272s & -18$^{\circ}$12'51.9579" &         14.9 &          4.8 &       47.9 &       49.2 & * & This paper \\
  2011-06-04 05:42:00.0 & 18h27m53.81859s & -18$^{\circ}$45'30.8046" &         15.6 &          9.3 &       71.7 &        7.1 & * & This paper\\
  2011-06-23 11:23:48.2 & 18h25m55.47963s & -18$^{\circ}$48'06.9712" &         16.1 &          5.5 &       73.2 &        0.2 & * & \citet{gul15} \\
  2012-07-18 04:13:00.0 & 18h32m14.67647s & -19$^{\circ}$24'19.3554" &         16.9 &          7.7 &       55.2 &      -76.0 & * & This paper \\
  2013-05-04 08:21:41.8 & 18h47m52.53356s & -19$^{\circ}$41'24.4265" &         18.7 &          8.4 &      -74.6 &       47.9 & * & \citet{olk15} \\
  2013-05-04 08:22:00.0 & 18h47m52.53305s & -19$^{\circ}$41'24.4265" &         19.3 &          9.2 &      -74.6 &       48.0 & * & This paper \\
  2014-07-23 14:25:59.1 & 18h49m31.74100s & -20$^{\circ}$22'23.9915" &         30.4 &          3.7 &       -7.5 &      -79.7 &   & \citet{pas16} \\
  2014-07-23 14:25:59.1 & 18h49m31.74048s & -20$^{\circ}$22'23.9502" &         23.0 &         44.9 &       -7.5 &      -79.7 &   & \citet{pas16}  \\
  2014-07-24 11:42:20.0 & 18h49m26.51393s & -20$^{\circ}$22'37.1172" &         11.3 &        -14.6 &      -65.8 &      -28.7 &   & \citet{pas16}  \\
  2014-07-24 11:42:20.0 & 18h49m26.51337s & -20$^{\circ}$22'37.0734" &          3.4 &         29.1 &      -65.8 &      -28.7 &   & \citet{pas16}  \\
  2015-06-29 16:02:00.0 & 19h00m49.70680s & -20$^{\circ}$41'40.4308" &         22.8 &         10.7 &      -41.9 &       80.3 & * & This paper \\
  2015-06-29 16:54:41.4 & 19h00m49.47778s & -20$^{\circ}$41'40.9707" &         22.1 &         12.7 &      -39.4 &       81.2 & * & \citet{pas17} \\
  2016-07-19 20:53:45.0 & 19h07m22.10999s & -21$^{\circ}$10'28.2320" &         24.1 &         11.6 &       56.5 &      -71.7 & * & This paper \\
\hline
\end{tabular}
\tablefoot{A flag * is indicated if the position was used in the NIMAv8 ephemeris (see Sec.~\ref{S:NIMA}).}
\label{T:Pluto-positions}
\end{center}
\end{table*}

\subsection{Astrometric positions from other publications}\label{Ss:astrometryOther}

Several authors have published circumstances of an occultation by Pluto \citep[e.g.][]{mil93,sic03,ell03,you08,per08,gul15,olk15,pas16,pas17}. 
From these circumstances (coordinates of the observer, mid-time of the occultation and impact parameter), it is possible to derive an offset between the observation deduced from these circumstances and a reference ephemeris. The procedure, based on the Bessel method used to predict stellar occultations, is described in Appendix~\ref{A:method} and the details of computation for each occultation are presented in Appendix~\ref{A:occ}. The Pluto's positions deduced from occultations published in other articles besides those of \cite{mez18} are presented in Table~\ref{T:Pluto-positions}.

The positions derived from \cite{pas16} involving single chord events and faint occulted stars, are not accurate enough to discriminate North and South solutions, i.e. to decide if Pluto's centre as seen from the observing site passed North or South of the star. Finally, these positions were not used in the orbit determination.

\section{NIMA ephemeris of Pluto}\label{S:NIMA} 

NIMA (Numerical Integration of the Motion of an Asteroid) has been developed in order to refine the  orbits of small bodies, in particular TNOs and Centaurs studied by stellar occultations technique \citep{des15}.  It consists of the numerical integration of the equations of motion perturbed by gravitational accelerations of the planets (Mercury to Neptune). The Earth and Moon are considered at their barycentre and the masses and the positions of the planets are from JPL DE436. 

The use of other masses and positions for planetary ephemeris produces insignificant changes, for example, the difference between the solution using DE436 and the solution using INPOP17a \citep{INPOP17a} for Pluto, is less than 0.06~mas on the 1985-2025 period. Moreover, there is no need to take into account the gravitational perturbations of the biggest TNOs. For example, by adding the 6 biggest TNOs (Eris, Haumea, 2007 OR10, Makemake, Quaoar and Sedna) in the model, the difference between the solutions with and without the biggest TNOs are about 0.04~mas in right ascension and declination on the 1985-2025 period, which is 100 times smaller than the mas-level accuracy of the astrometric positions.

The state vector (the heliocentric vector of position and velocity of the body at a specific epoch)  is refined by fitting to astrometric observations with the least square method.  The main advantage of NIMA is allowing for the use of observations published on the Minor Planet Center\footnote{The Minor Planet Center is in charge of providing astrometric measurements,  orbital elements of the solar system small bodies : \url{http://minorplanetcenter.net}.} together with unpublished observations or astrometric positions of occultations. The quality of the observations is taken into
 account with a specific weighting scheme, in particular, it takes advantages of the high accuracy of occultations. Finally, after fitting to the observations, NIMA can provide an ephemeris through a bsp file format readable by the SPICE library\footnote{The SPICE Toolkit is a library developed by NASA dedicated to space navigation and providing in  particular a list of routines related to ephemeris: \url{http://naif.jpl.nasa.gov/naif/index.html}.}.

As NIMA is representing the motion of the centre of mass of an object, it allows to compute the position of the Pluto's system barycentre and not the position of Pluto's centre itself. To deal with positions derived from occultations, we need an additional ephemeris representing the position of Pluto relative to its system barycentre. For that purpose, we use the most recent ephemeris PLU055 developed in 2015. The occultation-derived positions are then corrected from the offset between Pluto and the Pluto's system barycentre (see Table~\ref{T:Pluto-positions}) to derive the barycentric positions from the occultations, then used in the NIMA fitting procedure. 

The precisions of the positions in right ascension and in declination derived from the occultations are provided in Table~\ref{T:offsetEM} for occultations presented in \cite{mez18} and in Appendix~\ref{A:occ} for other publications. This precision is deduced from a specific model and reduction (for occultations in \citealt{mez18}) and from the precisions of timing and impact parameters (for other publications),  without any estimation of systematic errors. For a realistic estimation of the orbit accuracy, the weighting scheme in the orbit fit needs to take into account the systematic errors (see \citealt{des15} for details). The global accuracy for the positions used in the fitting depends on the accuracy of the stellar positions (from 0.1 to 2~mas), the precision of the derived position (from 0.1~mas to 11~mas), and the accuracy of the Pluto body-Pluto system barycentre ephemeris (estimated to 1-5~mas).

The errors on Pluto's centre determination have in fact various sources: the noise present in each occultation light curve and the spatial distribution of the occultation chords across the body. Assuming a normal noise, a formal error on the centre of the planet can then be derived, using a  classical least-square fitting and $\chi^2$ estimation. However, other systematic errors may also be present, such as problems in the absolute timing registration, slow sky transparency variations that make the photometric noise non-gaussian. Finally, the particular choice of the atmospheric model may also induce systematic biases in the centre determination. All those systematic errors are difficult to trace back. In that context, it is instructive to compare the reconstructions of the  geometry of a given occultation by independent groups that used different chords and different Pluto's atmospheric models. For example, occultations on 21 August 2002, 4 May 2013 and 29 June 2015 (see Table~\ref{tab_Pluto_RMS_NIMA}) indicate differences of few mas, which is much higher than the respective internal precisions (order of 0.1~mas). Case by case studies should be undertaken to explain those inconsistencies. This remains out of the scope of this paper. Meanwhile, for the weighting scheme in the orbit fit, we adopt the estimated precision presented in Table~\ref{tab_Pluto_RMS_NIMA} taking into account an estimation of systematic errors for each occultation.

\begin{figure}
\centering
\includegraphics[width=1.0\columnwidth]{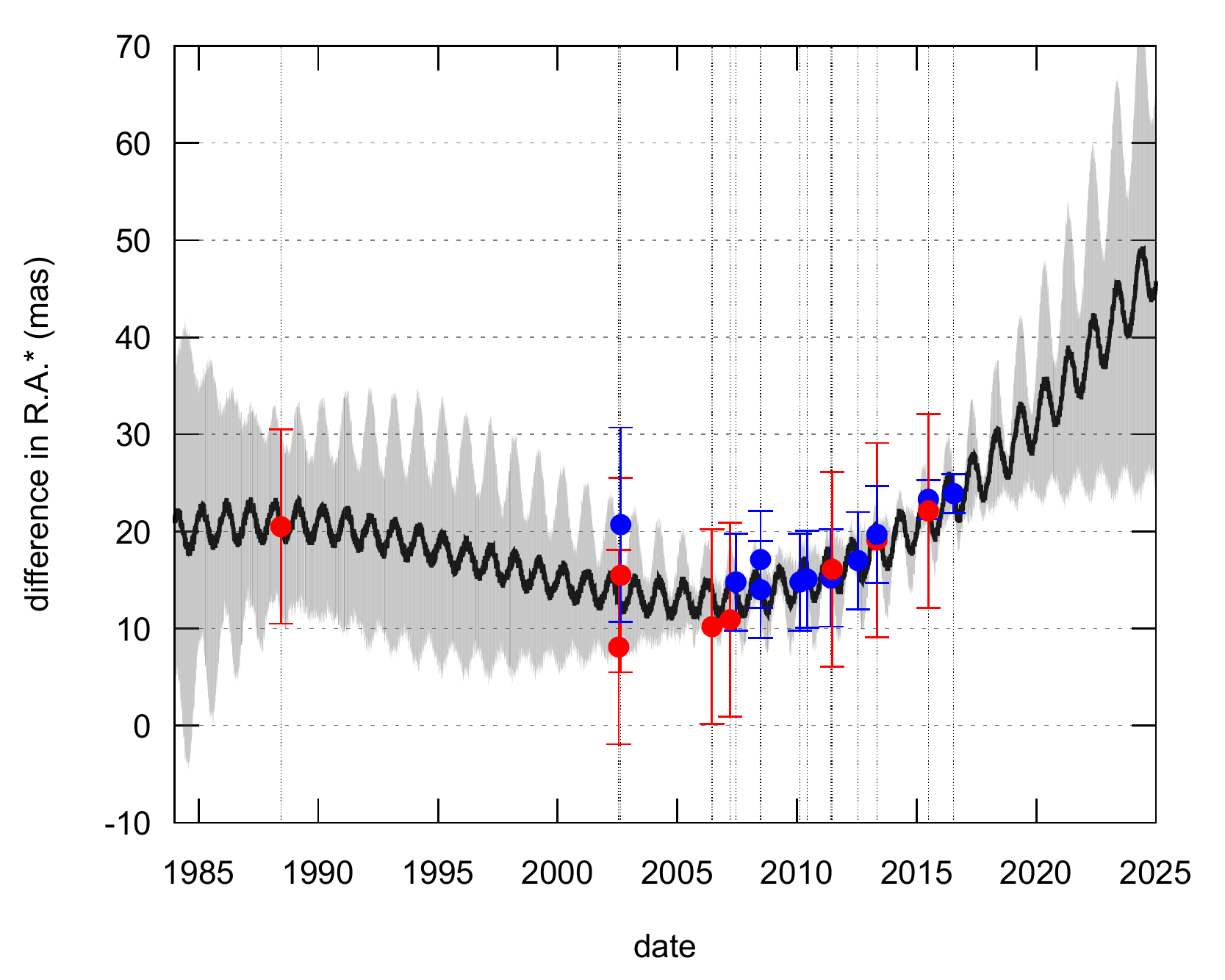}  
\includegraphics[width=1.0\columnwidth]{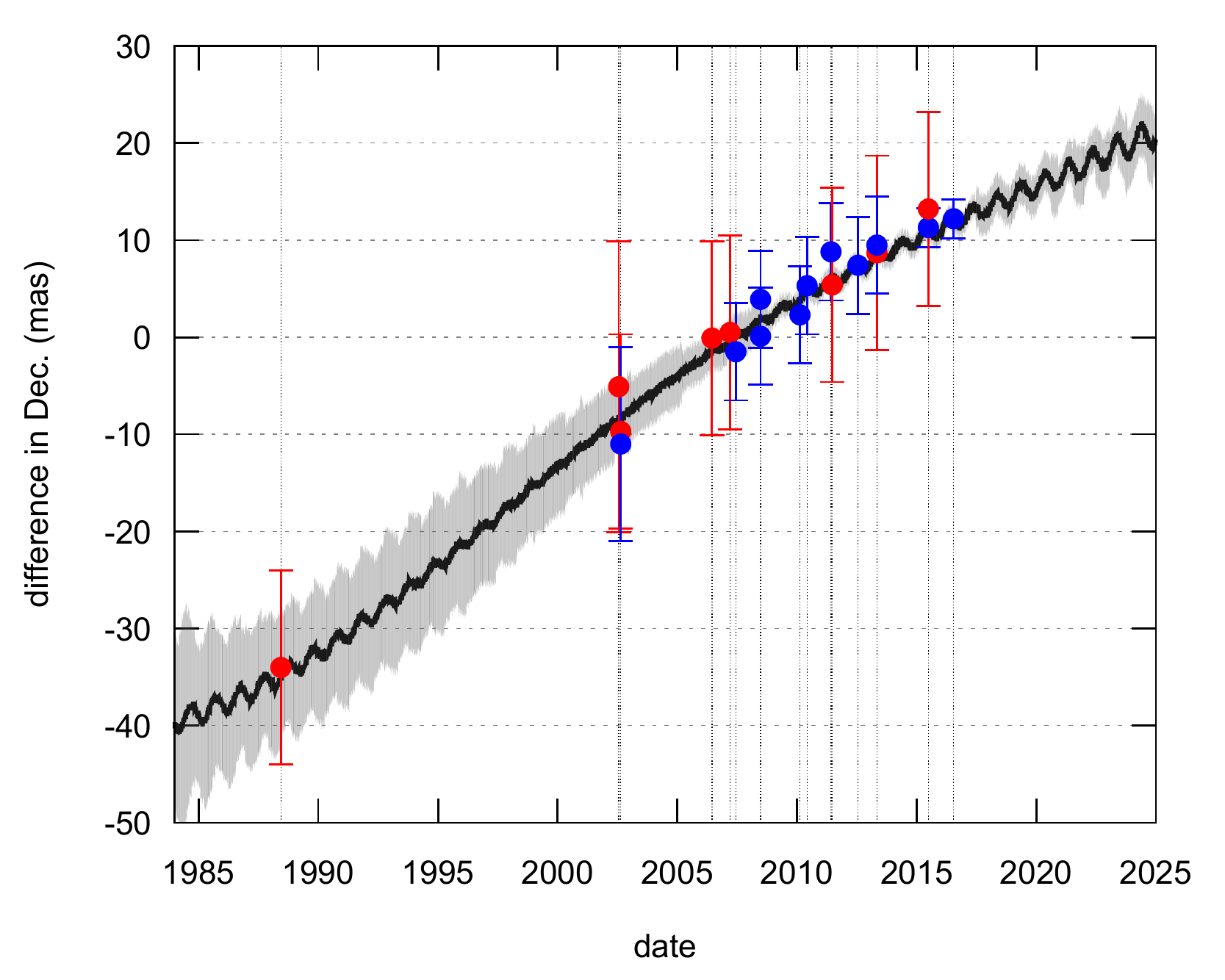}
\caption[Difference between NIMAv8 and JPL DE436 ephemeris of Pluto's system barycentre]{Difference between NIMAv8 and JPL DE436 ephemeris of Pluto's system barycentre (black line) in right ascension (weighted by $\cos \delta$) and in declination. Blue bullets and their estimated precision in error bar represent the positions coming from the occultations studied in this work and red bullets represent the positions deduced from other publications. The gray area represents the 1$\sigma$ uncertainty of the NIMA orbit. Vertical gray lines indicate the date of the position for a better reading on the $x$-axis. Note that the angular diameter of Pluto, as seen from Earth, is about 115~mas, while the atmosphere detectable using ground-based stellar occultations subtends about 150~mas on the sky.}
\label{NIMAv8}
\end{figure}

Figure~\ref{NIMAv8} shows the difference between NIMA\footnote{The NIMAv8 ephemeris is available on \url{http://lesia.obspm.fr/lucky-star/nima.php}.} and JPLDE436 ephemeris of Pluto's barycentre in right ascension (weighted by $\cos \delta$) and declination. The blue bullets and the error bars represent the positions and their estimated precision from our occultations, the red bullets represent the positions from occultations not listed in \cite{mez18}:  \cite{mil93,sic03,ell03,you08,per08,gul15,olk15,pas17}, and the gray area represents the one sigma uncertainty of the NIMAv8 ephemeris. 

\begin{figure}
\centering
\includegraphics[width=1.0\columnwidth]{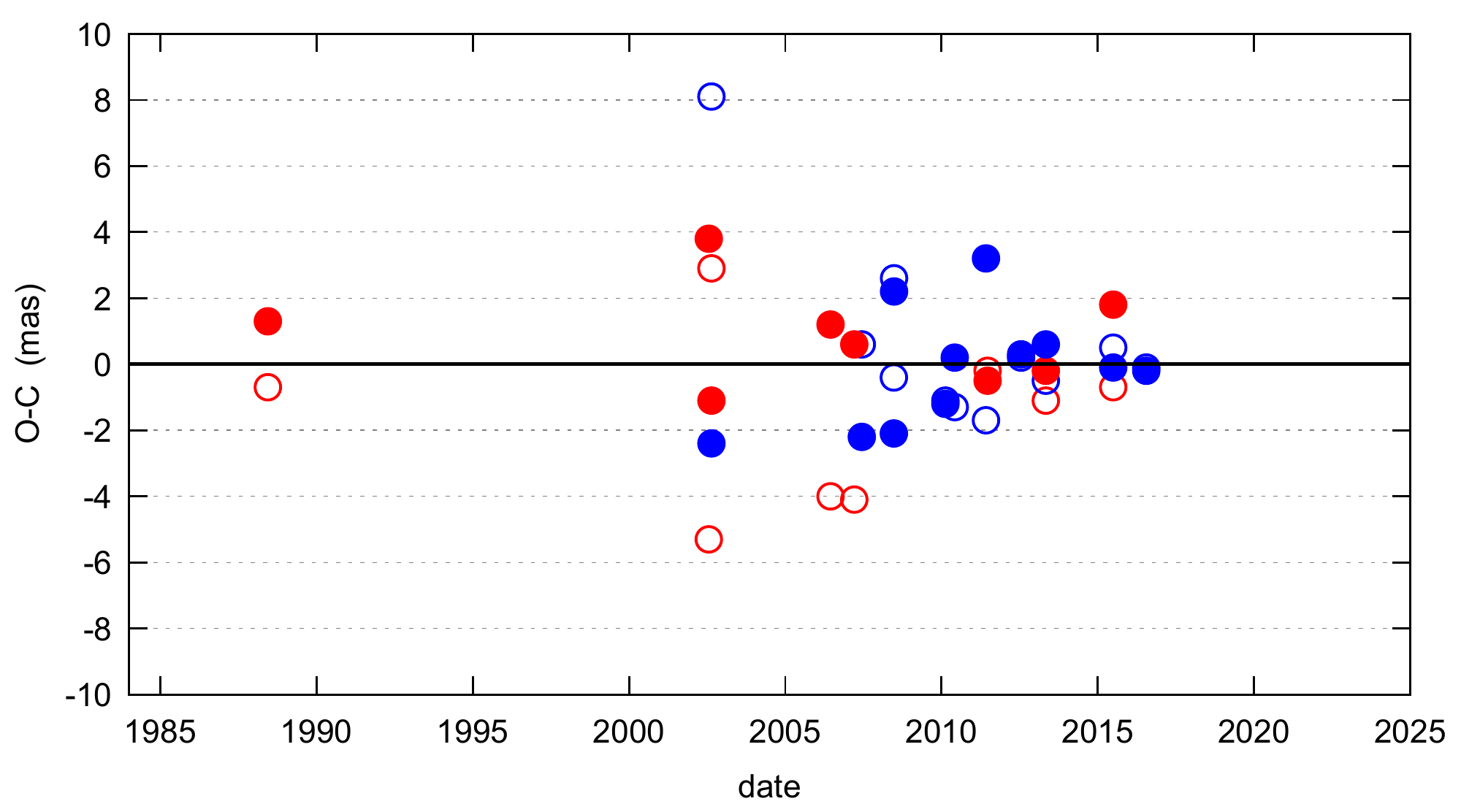}  
\caption[Residuals of Pluto's system barycentre positions compared to NIMAv8]{Residuals of Pluto's system barycentre positions compared to NIMAv8. Circles are for right ascension weighted by $\cos \delta$ and bullets are for declination. Blue color is used for the positions coming from the occultations studied in this work and red color is for the positions deduced from other publications}
\label{F:omcnima}
\end{figure}

Table~\ref{tab_Pluto_RMS_NIMA} and Fig.~\ref{F:omcnima} provide the residuals (O-C) of the positions derived from the occultations, compared with the NIMAv8 ephemeris, and the estimated precision of the positions used in the weighting scheme. After 2011, residuals are mostly below the mas level, which is much better than any ground-based astrometric observation of Pluto. In that context, other classical observations of Pluto, such as CCD, appear to be less useful for ephemerides of Pluto during the period covered by the occultations 1988-2016.

\begin{table}
\caption{Residuals (O-C) related to NIMAv8 ephemeris of Pluto system barycentre. Estimated precision in mas in right ascension and declination used for the fit is also indicated.}
\label{tab_Pluto_RMS_NIMA}
\centering
\begin{tabular}{lrrrr}
\hline
date & $\Delta\alpha\cos(\delta)$ & $\Delta\delta$  & $\sigma_\alpha$ & $\sigma_\delta$ \\ 
(UTC) & (mas) & (mas) & (mas) & (mas)\\ 
\hline
1988-06-09 &  -0.7   &    1.3  &    10.0   &    10.0 \\  
2002-07-20 &  -5.3   &    3.8  &    10.0  &     15.0 \\ 
2002-08-21\tablefootmark{1} &   2.9   &   -1.1 &   10.0  &     10.0 \\ 
2002-08-21 &   8.1   &   -2.4 &       10.0  &   10.0 \\  
2006-06-12 &  -4.0   &    1.2 &       10.0  &   10.0 \\  
2007-03-18 &  -4.1   &    0.6 &       10.0  &   10.0 \\  
2007-06-14 &   0.6   &   -2.2 &     5.0  &   5.0 \\  
2008-06-22 &  -0.4   &   -2.1 &     5.0  &   5.0 \\  
2008-06-24 &   2.6   &    2.2 &     5.0  &   5.0 \\  
2010-02-14 &  -1.1   &   -1.2 &     5.0  &   5.0 \\  
2010-06-04 &  -1.3   &    0.2 &     5.0  &   5.0 \\  
2011-06-04 &  -1.7   &    3.2 &     5.0  &   5.0 \\  
2011-06-23 &  -0.2   &   -0.5 &     10.0  &   10.0 \\  
2012-07-18 &   0.2   &    0.3 &     5.0  &   5.0 \\  
2013-05-04\tablefootmark{2} &  -1.1   &   -0.2 &     10.0  &   10.0 \\  
2013-05-04 &  -0.5   &    0.6 &     5.0  &   5.0 \\  
2015-06-29 &   0.5   &   -0.1 &     2.0  &   2.0 \\  
2015-06-29\tablefootmark{3} &  -0.7   &    1.8 &     10.0  &   10.0 \\  
2016-07-19 &  -0.1   &   -0.2 &     2.0  &   2.0 \\
\hline
\end{tabular}
\tablefoot{%
\tablefoottext{1}{Taken from \cite{ell03}.}
\tablefoottext{2}{Taken from \cite{olk15}.}
\tablefoottext{3}{Taken from \cite{pas17}.}
}
\end{table}

\begin{figure}
\centering
\includegraphics[width=1.0\columnwidth]{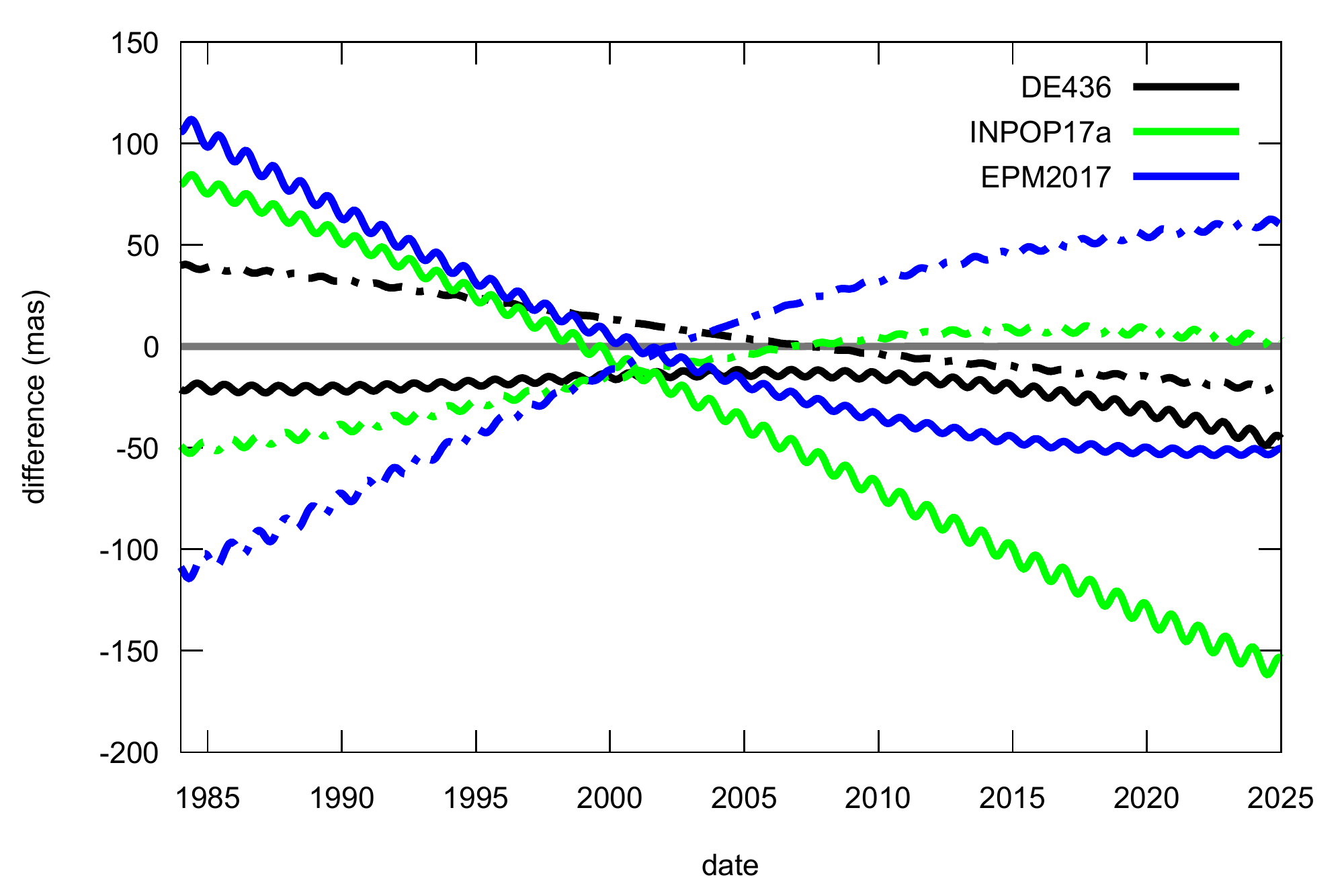}  
\caption[Difference between several ephemerides]{Difference in right ascension weighted by $\cos \delta$ (solid line) and declination (dotted line) between several ephemerides of Pluto system barycentre: JPL DE436, INPOP17 and EPM2017, compared to NIMAv8.}
\label{F:diffephem}
\end{figure}

Figure~\ref{F:diffephem} shows the difference in right ascension and declination between the most recent ephemerides of Pluto system barycentre: JPL DE436, INPOP17a \citep{INPOP17a} and EPM2017 \citep{pit14} compared to NIMAv8. These differences are mostly due to data and weights used for the orbit determination. They reveal periodic terms in the orbit of Pluto system barycentre that are differently estimated in orbit determination. As described in \cite{des15}, the one-year period corresponds to the parallax induced by different geocentric distances given by the ephemerides. It is also another good indication of the improvement of the NIMAv8 ephemeris since the differences between these ephemerides reach 50-100~mas whereas the estimated precision of NIMAv8 is 2-20~mas on the same period.

\section{Discussion}\label{S:discussion} 

The NIMA ephemeris allows very accurate predictions of stellar occultation by Pluto in the forthcoming years within a few mas level. In particular, we have predicted an occultation of a magnitude 13 star\footnote{The star position in Gaia DR2 at the epoch of the occultation is 19h22m10.4687s in right ascension and  $-21^{\circ}58'49.020"$ in declination.} by Pluto on August 15, 2018, above North America to the precision of 2.5~mas, representing only 60~km on the shadow path and a precision of 4~s in time. As shown in \cite{mez18}, the observation of a central flash allows to probe the deepest layers of Pluto's atmosphere. The central flash can be observed in an small band about 50~km around the centrality path. By reaching a precision of tens of km, we were able to gather observing stations along the centrality and to highly increase the probability of observing a central flash. 

\begin{figure}
\centering
\includegraphics[width=1.0\columnwidth]{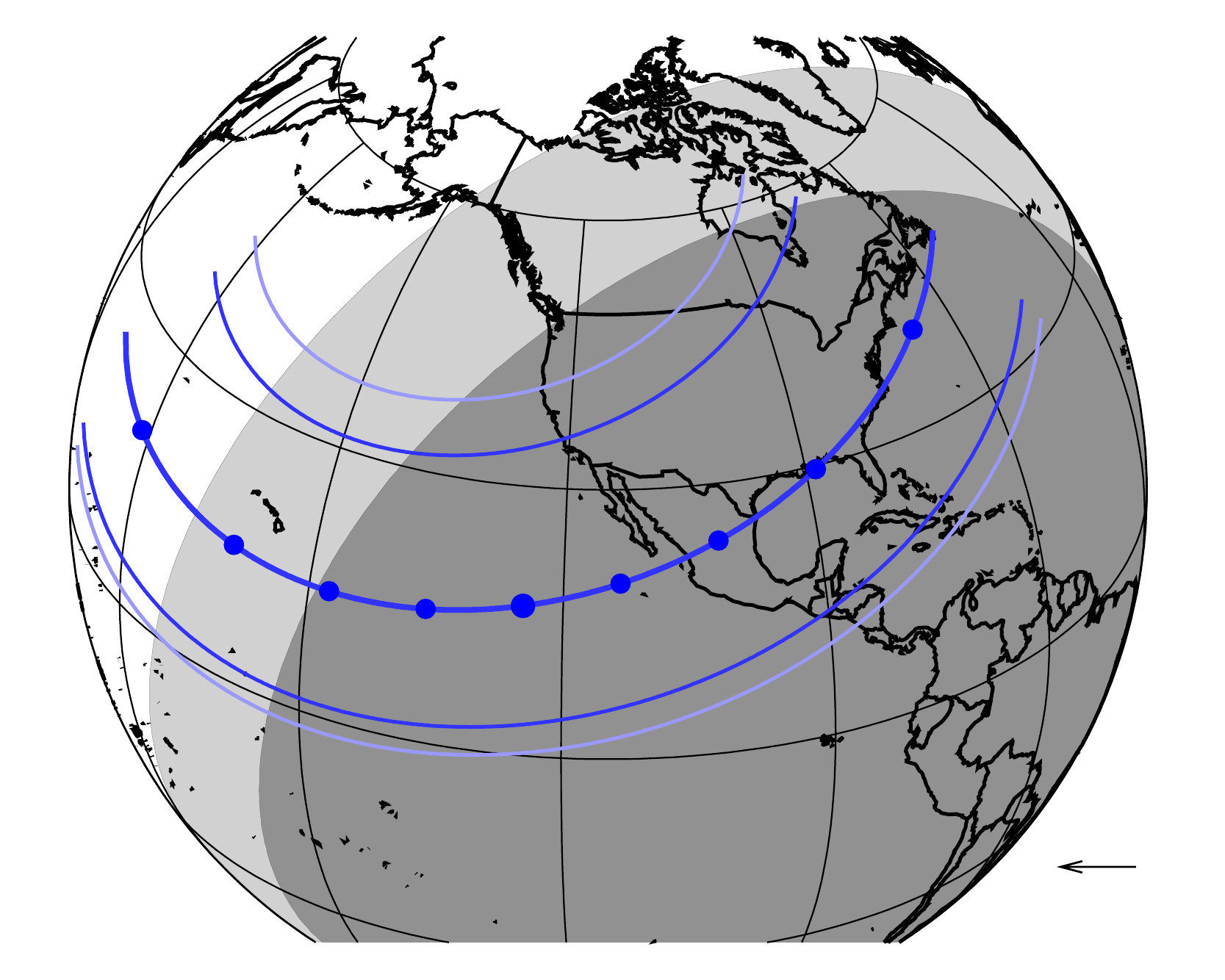} \\
\includegraphics[width=1.0\columnwidth]{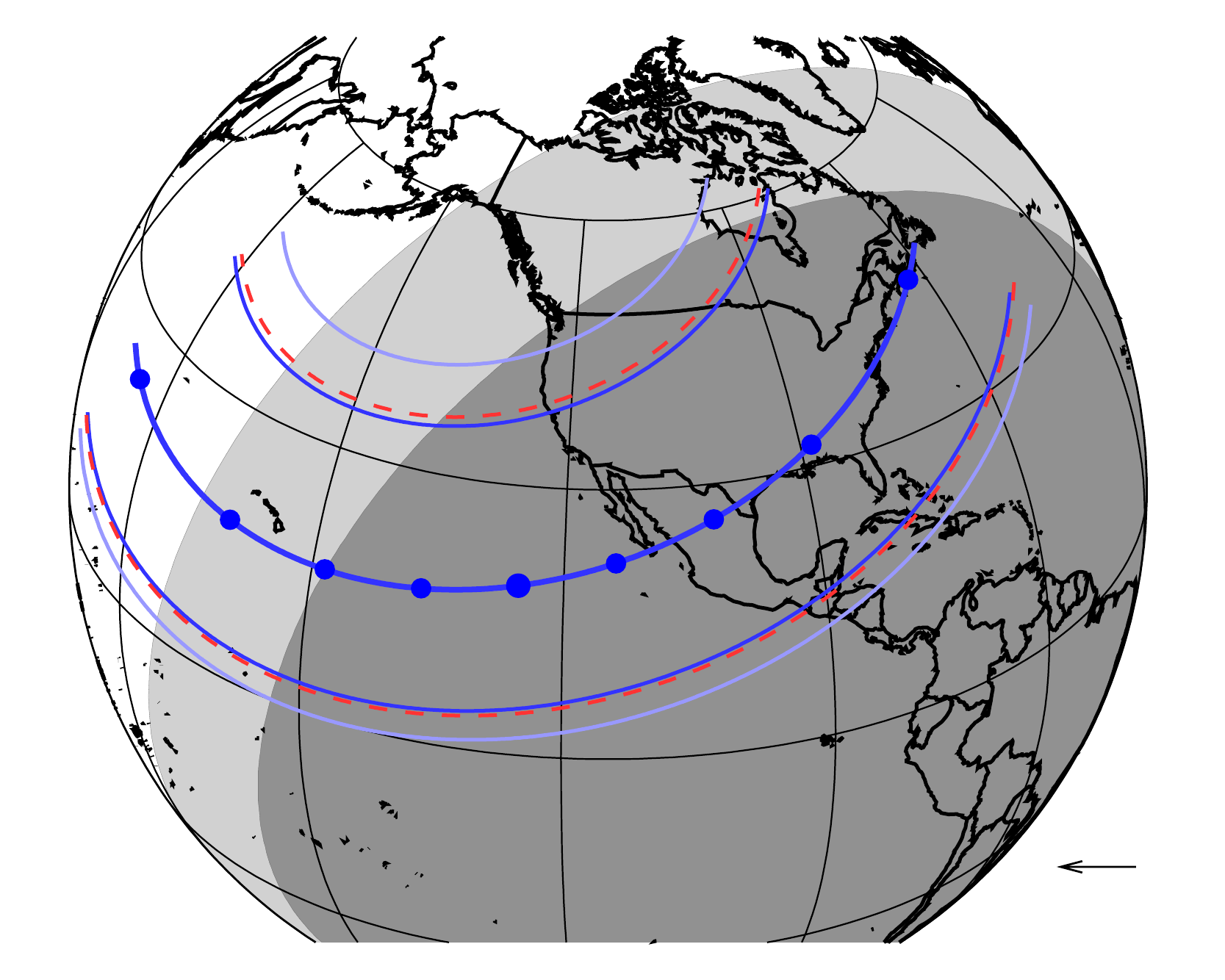}
\caption[Prediction of the occultation by Pluto on 15 August 2018]{
Prediction of the occultation by Pluto on 15  August 2018, using JPL DE436/PLU055 (top) and NIMAv8/PLU055 (bottom) ephemerides. The red dashed lines represent the 1$\sigma$ uncertainty on the path, taking into account the uncertainties of NIMAv8 ephemeris and of the star position. The bullets on the shadow central line are plotted every minute. The dark and light blue thinner lines are the shadow limits corresponding the stellar half-light level and 1\% stellar drop level (the practical detection limit), respectively.}
\label{map2018}
\end{figure}

The prediction of the August 15, 2018 Pluto occultation was used to assess the accuracy of our predictions using the NIMA approach. 
Figure~\ref{map2018} represents the prediction of the occultation by Pluto on August 15, 2018 using two different ephemerides: JPL DE436/PLU055 and NIMAv8/PLU055. The prediction using JPL ephemerides is shifted by 36.8~s and 8~mas south (representing about 190~km) compared to the prediction with NIMAv8 ephemeris. Several stations detected the occultation, some of them revealing a central flash. For instance, observers at George Observatory (Texas, USA) report a central flash of typical amplitude 20\%, compared to the unocculted stellar flux (T. Blank and P. Maley, private communications).

As the amplitude of the flash roughly scales as the inverse of the closest approach (C/A) distance of the station to the shadow centre, the amplitude may serve to estimate the C/A distance. A central flash reported by \cite{sic16} was observed at a station in New Zealand during the June 29, 2015 occultation. It had an amplitude of 13\%, with C/A distance of 42~km. Thus, the flash observed at George Observatory provides an estimated C/A distance of 25~km for that station. This agrees with the value predicted by the NIMAv8/PLU055 ephemeris, to within 3 km, corresponding to 0.12 mas. This is fully consistent with, but smaller than our 2.5 mas error bar quoted above, possibly indicating an overestimation of our prediction errors.

The precision of our predictions remains at few mas up to 2025 (in particular in declination) and it is even more important since the apparent position of Pluto as seen from Earth is moving away from the Galactic centre, making occultations by Pluto more rare.

\begin{figure}
\centering
\includegraphics[width=1.0\columnwidth]{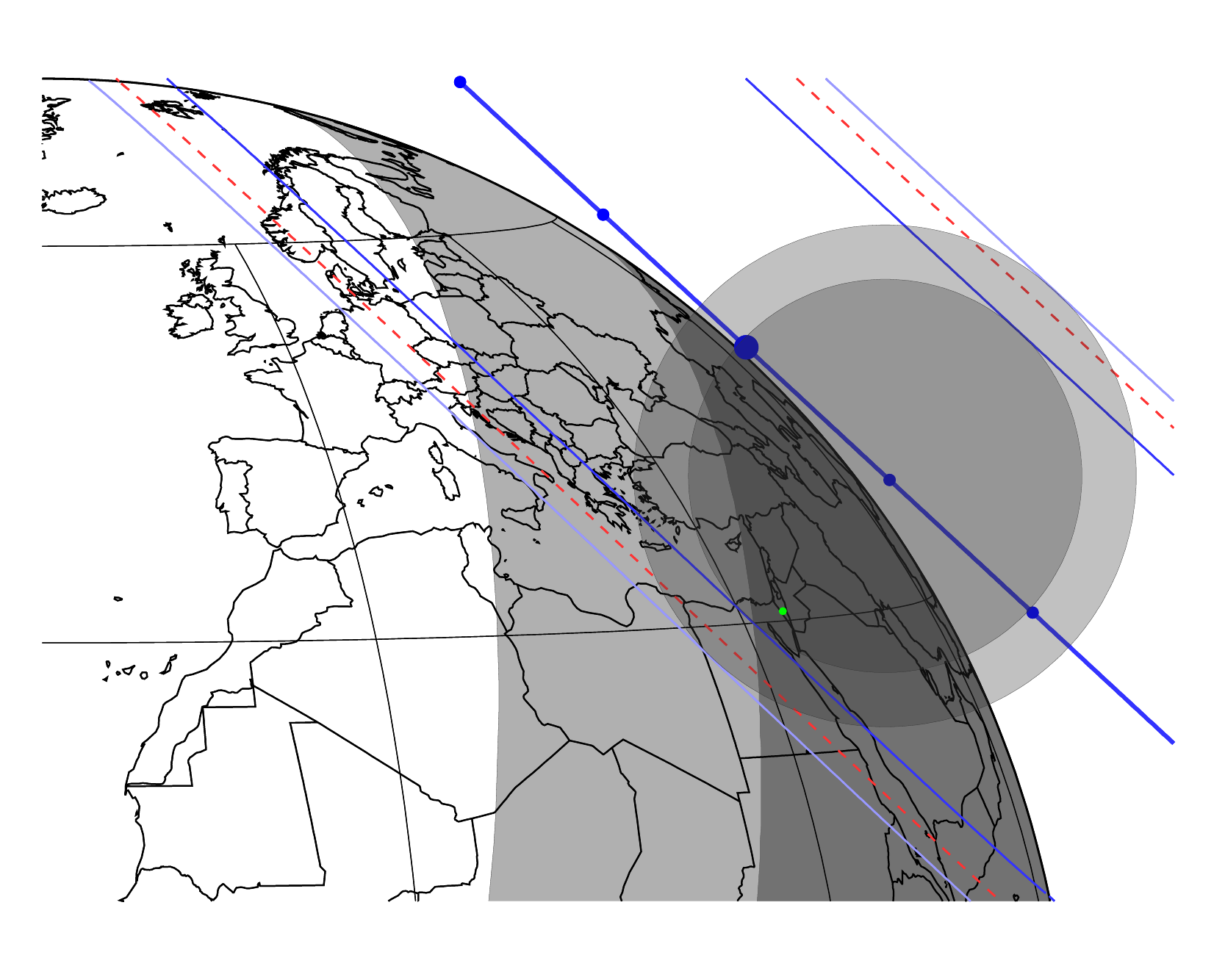} 
\caption[Postdiction of the occultation by Pluto on 19 August 1985]{
Postdiction of the Pluto's occultation of 19 August 1985,
using NIMAv8/PLU055 ephemerides. The shadow of Pluto at 17:59:54 (the mid-time of the occultation provided in \citealt{bro95}) is represented. The green bullet represents the WISE observatory. 
The red dashed lines represent the 1$\sigma$ uncertainty on the path. Areas in dark gray corresponds to full night (Sun elevation below -18$^\circ$) and areas in light grey corresponds to twilight (Sun elevation between -18$^\circ$ to 0$^\circ$), while day time regions are in white. The dark and light blue thinner lines are the shadow limits corresponding the stellar half-light level and 1\% stellar drop level (the practical detection limit), respectively.}
\label{map1985}
\end{figure}

Another point of interest is to look at past occultations. In particular, for the occultation of August 19, 1985, \cite{bro85} reported a single chord occultation of a magnitude 11.1 star\footnote{The star position in Gaia DR2 at the epoch of the occultation is 14h23m43.4575s in right ascension and $+03^{\circ}06'46.874"$ in declination.} by Pluto, showing a gradual shape possibly due to Pluto's atmosphere. The observation was performed at Wise observatory in Israel under poor conditions (low elevation, flares from passing planes, close to twilight). Thanks to Gaia DR2 providing the proper motion of the star and to NIMAv8, we make a postdiction of the occultation of August 19, 1985 (Fig.~\ref{map1985}). The nominal time for the occultation (the time of the closest approach between the geocentre and the centre of the shadow) is 17:58:57.1 (UTC) leading to a predicted mid-time of 17:59:49.8 (UTC) at Wise observatory. \cite{bro95} gave an approximate observed mid-time of the occultation for Wise observatory at 17:59:54 (about 4~s later than the prediction). The predicted shadow of Pluto at the same time is presented on the figure as well as the observatory's place as a green bullet. Taking into account the uncertainties of the NIMAv8 ephemeris and of the star position, the uncertainty in time for this occultation is about 20~s whereas the crosstrack uncertainty on the path is about 10~mas (representing 220~km). This is fully consistent with the fact that the occultation was indeed observed at Wise observatory.

\section{Conclusions}
Stellar occultations by Pluto provide accurate astrometric positions thanks to Gaia catalogues, in particular Gaia DR2.  We determine 18 astrometric positions of Pluto from 1988 to 2016 with an estimated precision of 2 to 10~mas.

These positions are used to compute an ephemeris of Pluto system's barycentre thanks to NIMA procedure with an unprecedented precision on the 1985-2015 period. This ephemeris NIMAv8 was used to study the possible occultation of Pluto observed in 1985 as well to predict the recent occultation by Pluto on August 15th, 2018 or the forthcoming occultations\footnote{See the predictions on the Lucky Star webpage \url{http://lesia.obspm.fr/lucky-star/predictions.php}.} with a precision of 2 mas, a result impossible to reach with classical astrometry and previous stellar catalogues. In fact, the presence of the usually unresolved Charon in classical images,
causes significant displacements of the photocentre of the system with respect
to its barycentre. As a consequence, and even modeling the effect of Charon, as in \cite{ben14}, accuracies below the 50 mas level are difficult to reach. 

This method can be extended, for instance for Chariklo, with an even better accuracy of the order of 1~mas \citep{des17}
and illustrates the power of stellar occultations not only for better studying those bodies,
but also for improving their orbital elements.

\begin{acknowledgements}
Part of the research leading to these results has received funding from the European Research Council under the European Community’s H2020 (2014–2020/ERC Grant Agreement No. 669416 “LUCKY STAR”).
This work has made use of data from the European Space Agency (ESA) mission
{\it Gaia} (\url{https://www.cosmos.esa.int/gaia}), processed by the {\it Gaia}
Data Processing and Analysis Consortium (DPAC,
\url{https://www.cosmos.esa.int/web/gaia/dpac/consortium}). Funding for the DPAC
has been provided by national institutions, in particular the institutions
participating in the {\it Gaia} Multilateral Agreement.
J.I.B.C. acknowledges CNPq grant 308150/2016-3.
M.A. thanks CNPq (Grants 427700/2018-3, 310683/2017-3 and 473002/2013-2) and FAPERJ (Grant E-26/111.488/2013).
G.B.R. is thankful for the support of the CAPES (203.173/2016) and FAPERJ/PAPDRJ (E26/200.464/2015-227833) grants.
This study was financed in part by the Coordenação de Aperfeiçoamento
de Pessoal de Nível Superior - Brasil (CAPES) - Finance Code 001. 
F.B.R.acknowledges CNPq grant 309578/2017-5.
A.R.G-J thanks FAPESP proc. 2018/11239-8.
R.V-M thanks grants: CNPq-304544/2017-5, 401903/2016-8, Faperj: PAPDRJ-45/2013 and E-26/203.026/2015 
P.S.-S. acknowledges financial support by the European Union’s Horizon 2020 Research and Innovation Programme, under Grant Agreement no 687378, as part of the project "Small Bodies Near and Far" (SBNAF).
\end{acknowledgements}

\bibliographystyle{aa}
\bibliography{references}

\begin{figure*}
\begin{subfigure}[b]{.5\linewidth}
\centering \includegraphics[width=0.9\columnwidth]{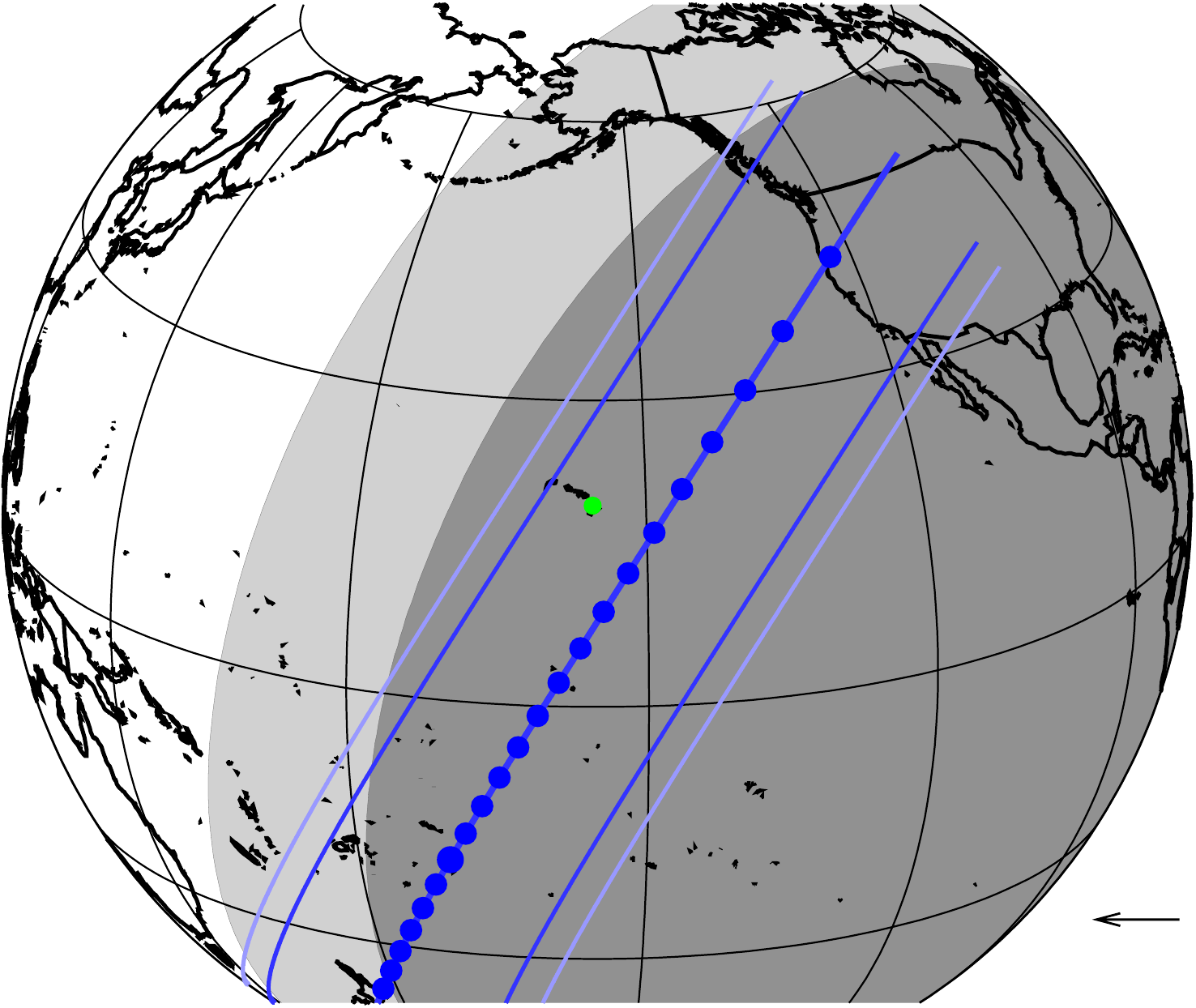}
\subcaption{2002-08-21}
\end{subfigure}%
\begin{subfigure}[b]{.5\linewidth}
\centering \includegraphics[width=0.9\columnwidth]{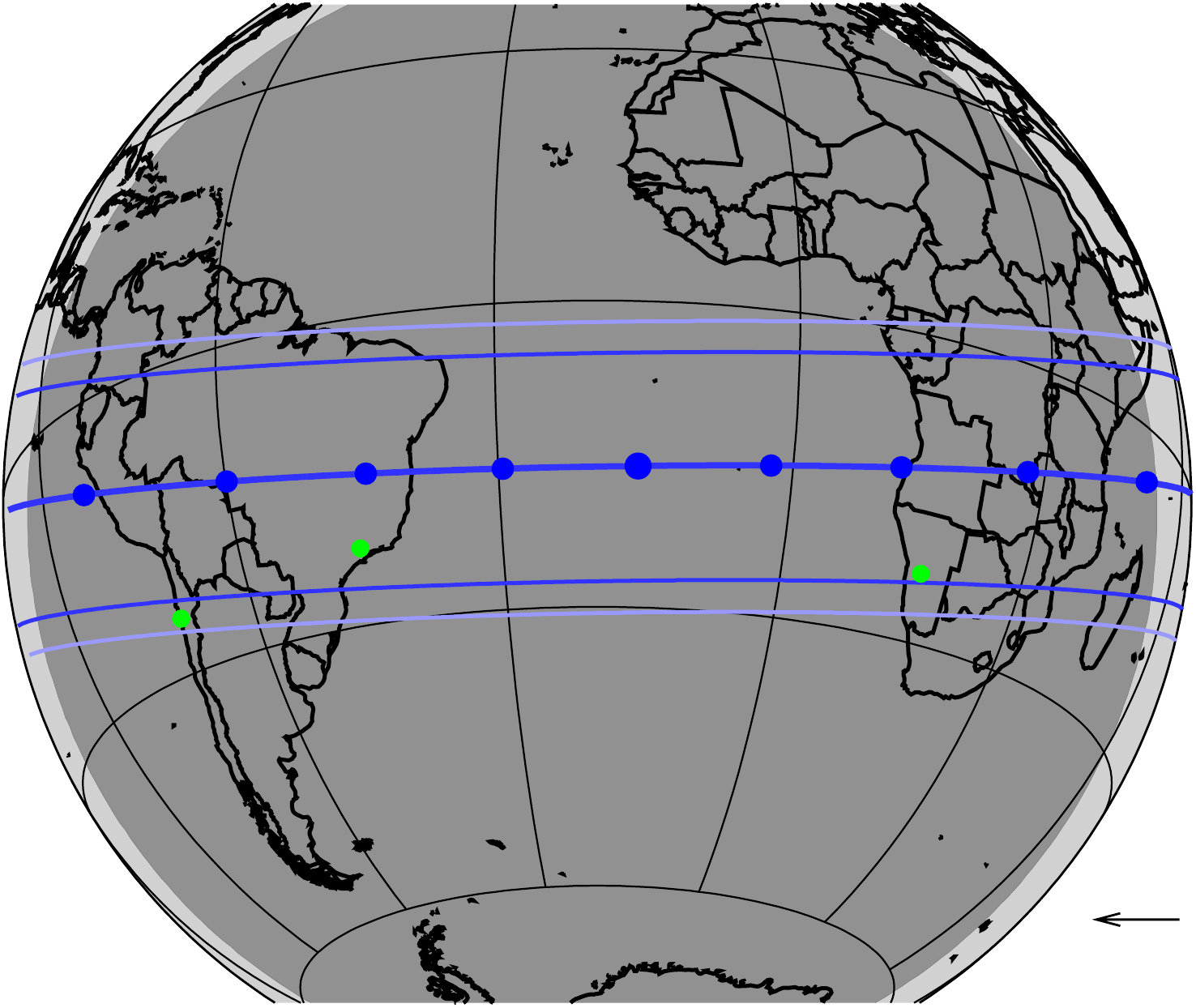}
\subcaption{2007-06-14}
\end{subfigure}%

\begin{subfigure}[b]{.5\linewidth}
\centering \includegraphics[width=0.9\columnwidth]{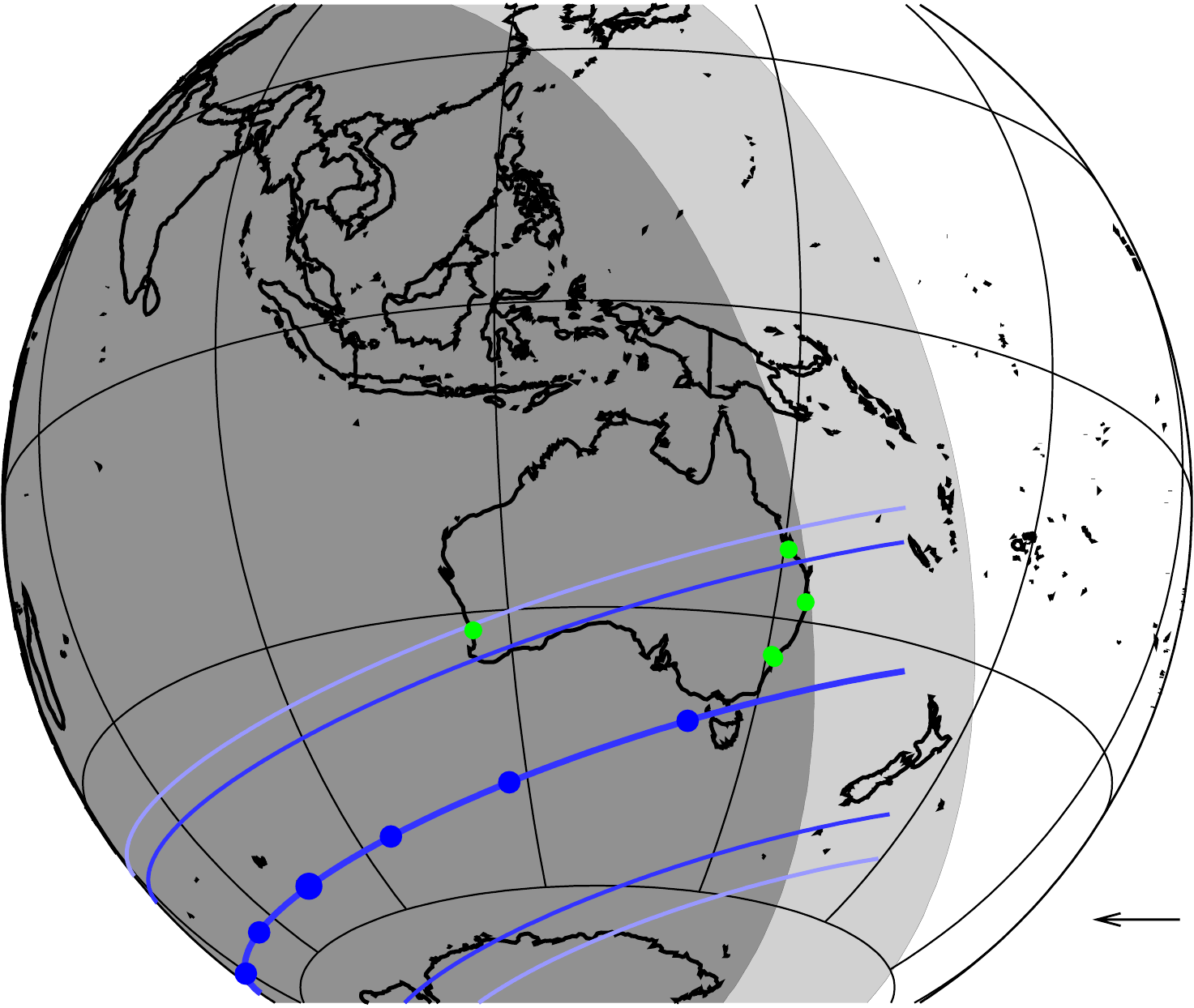}
\subcaption{2008-06-22}
\end{subfigure}%
\begin{subfigure}[b]{.5\linewidth}
\centering \includegraphics[width=0.9\columnwidth]{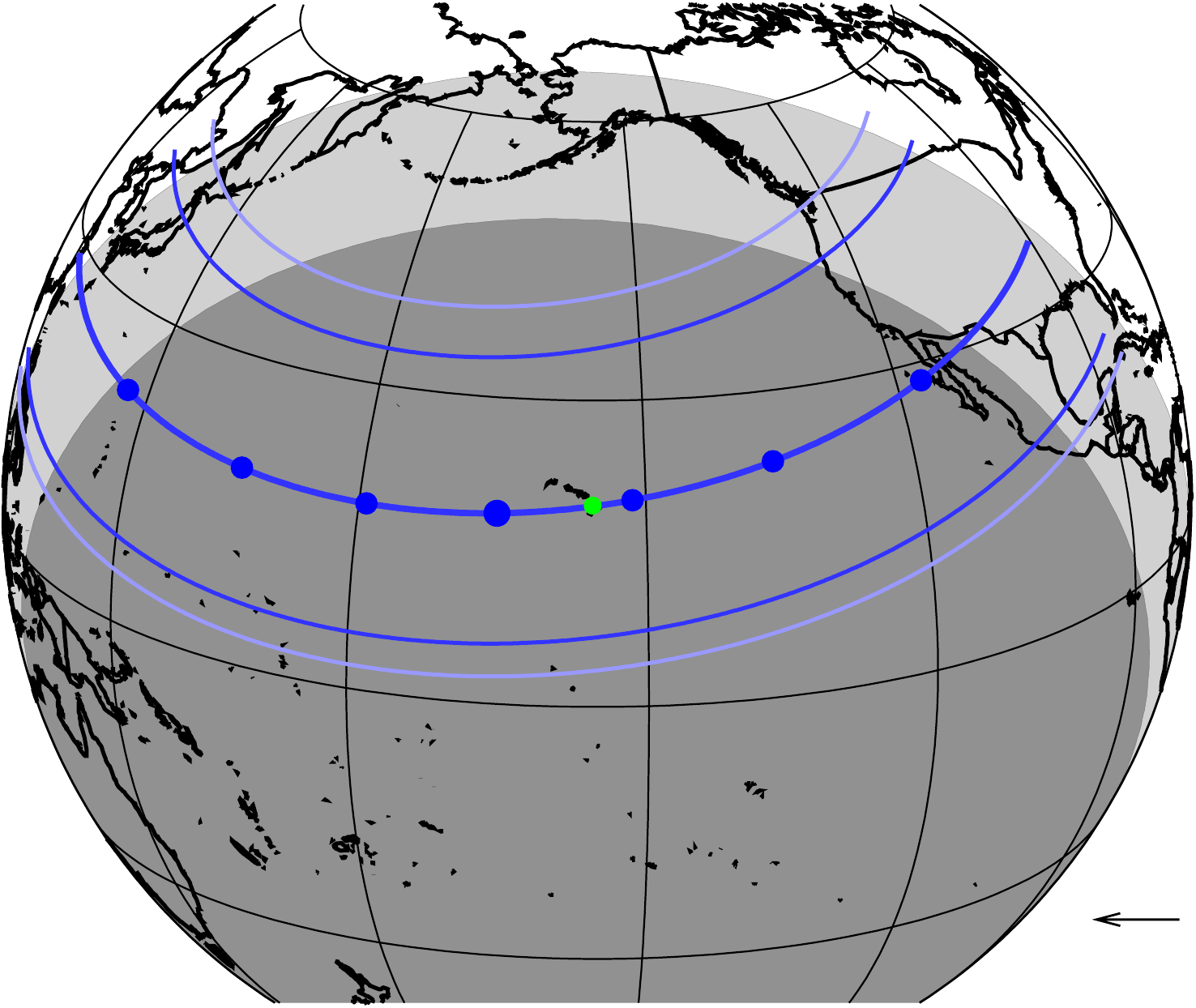}
\subcaption{2008-06-24}
\end{subfigure}%

\begin{subfigure}[b]{.5\linewidth}
\centering \includegraphics[width=0.9\columnwidth]{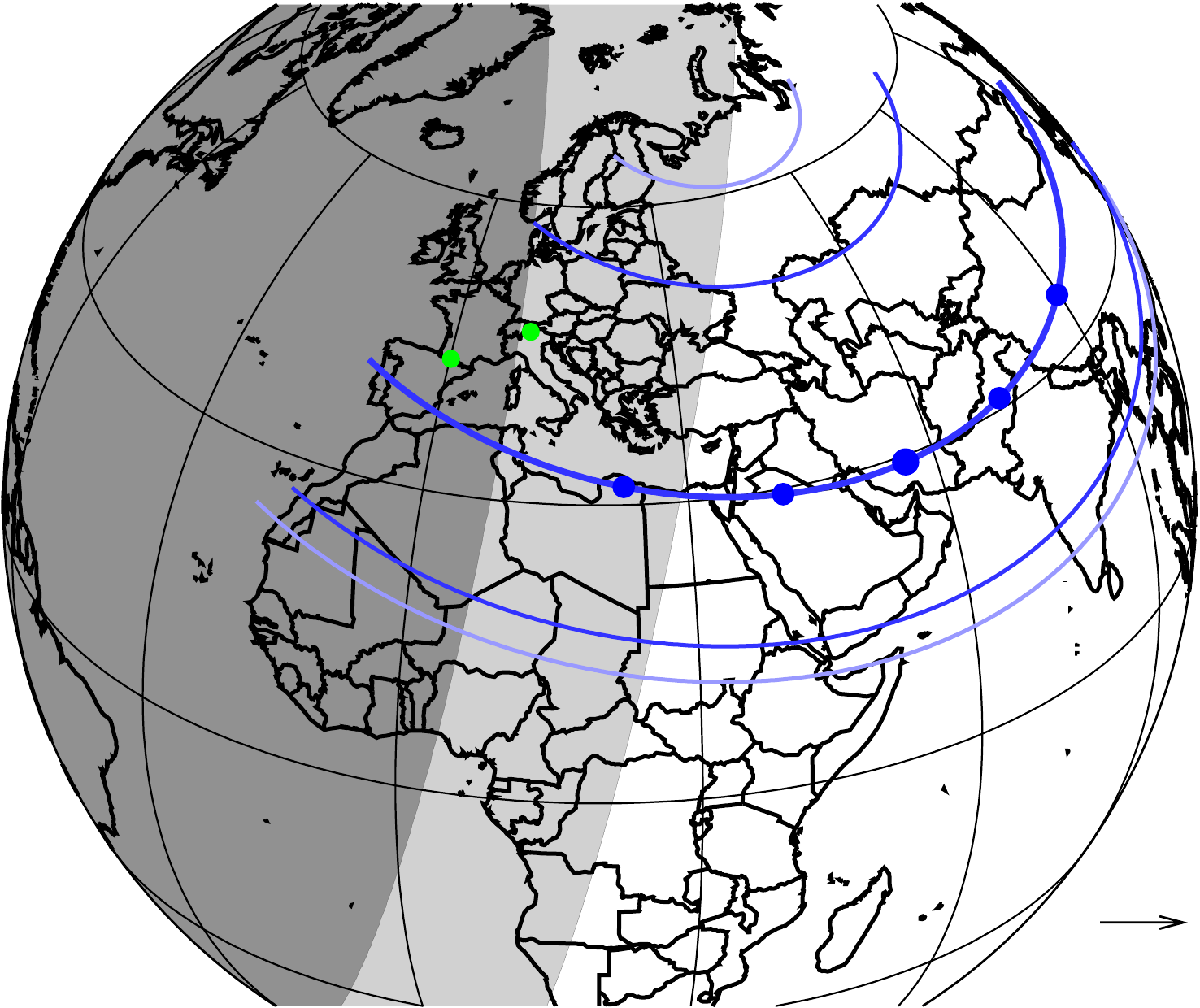}
\subcaption{2010-02-14}
\end{subfigure}%
\begin{subfigure}[b]{.5\linewidth}
\centering \includegraphics[width=0.9\columnwidth]{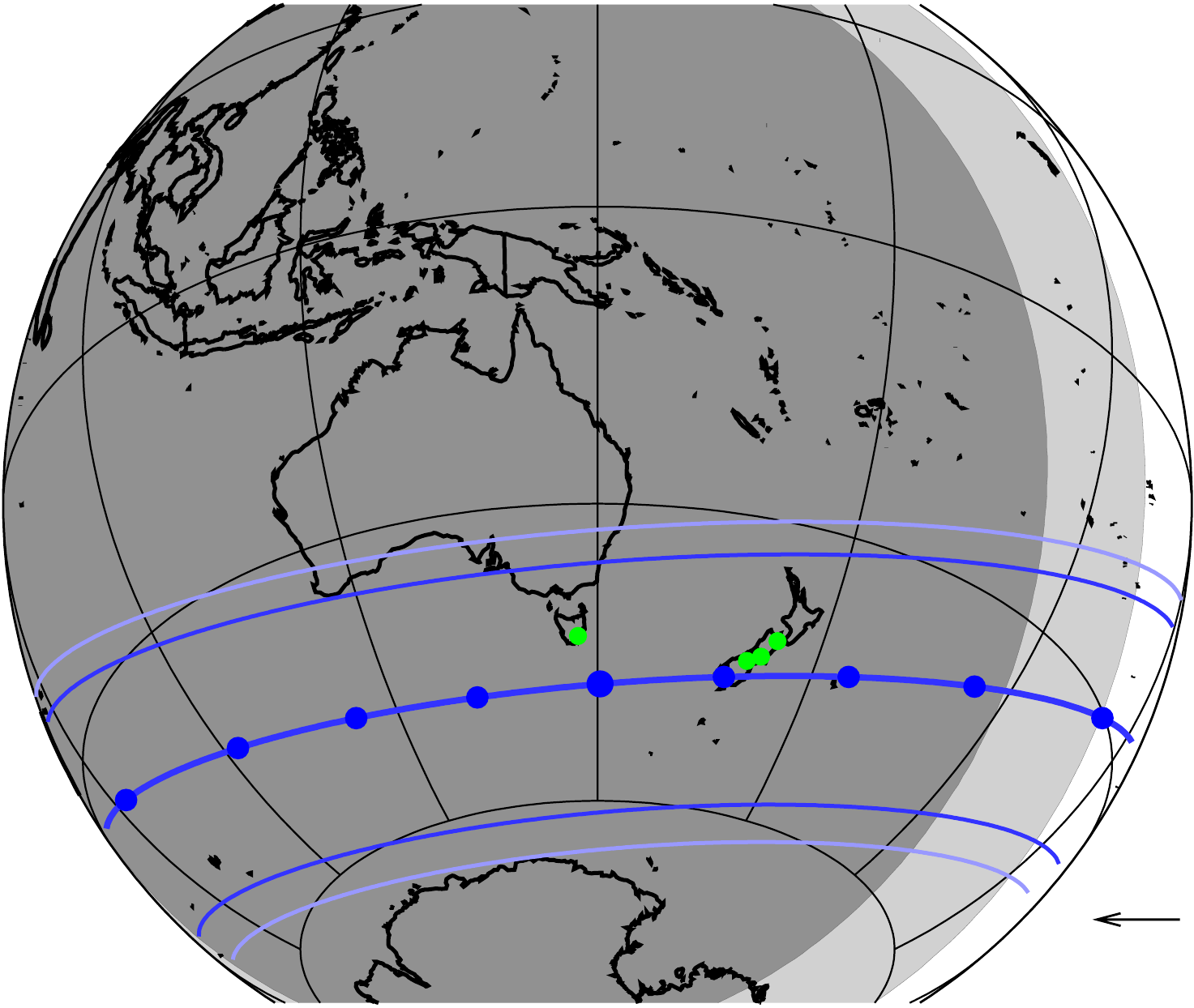}
\subcaption{2010-06-04}
\end{subfigure}%

\end{figure*}

\begin{figure*}[htb]\ContinuedFloat

\begin{subfigure}[b]{.5\linewidth}
\centering \includegraphics[width=0.9\columnwidth]{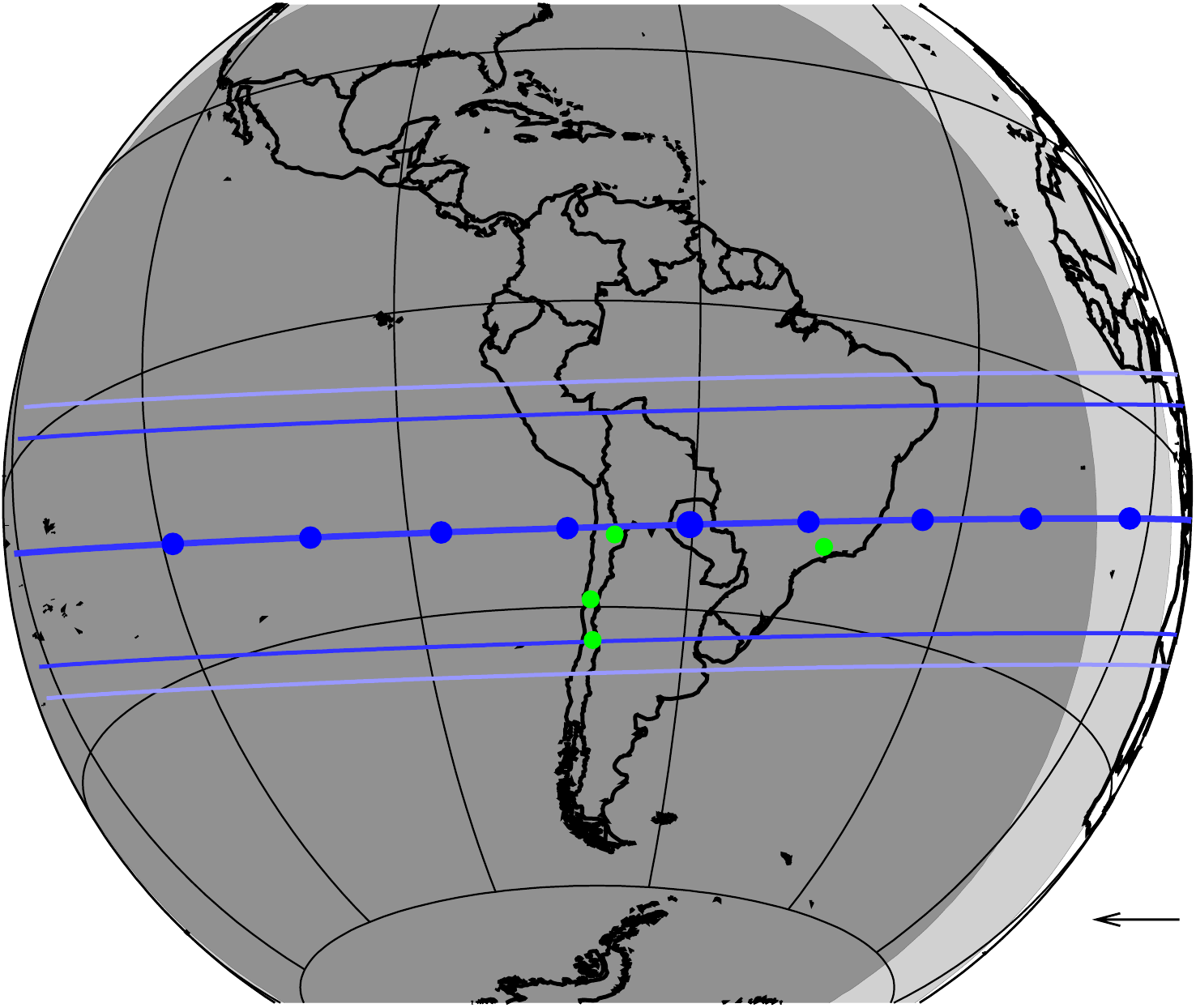}
\subcaption{2011-06-04}
\end{subfigure}%
\begin{subfigure}[b]{.5\linewidth}
\centering \includegraphics[width=0.9\columnwidth]{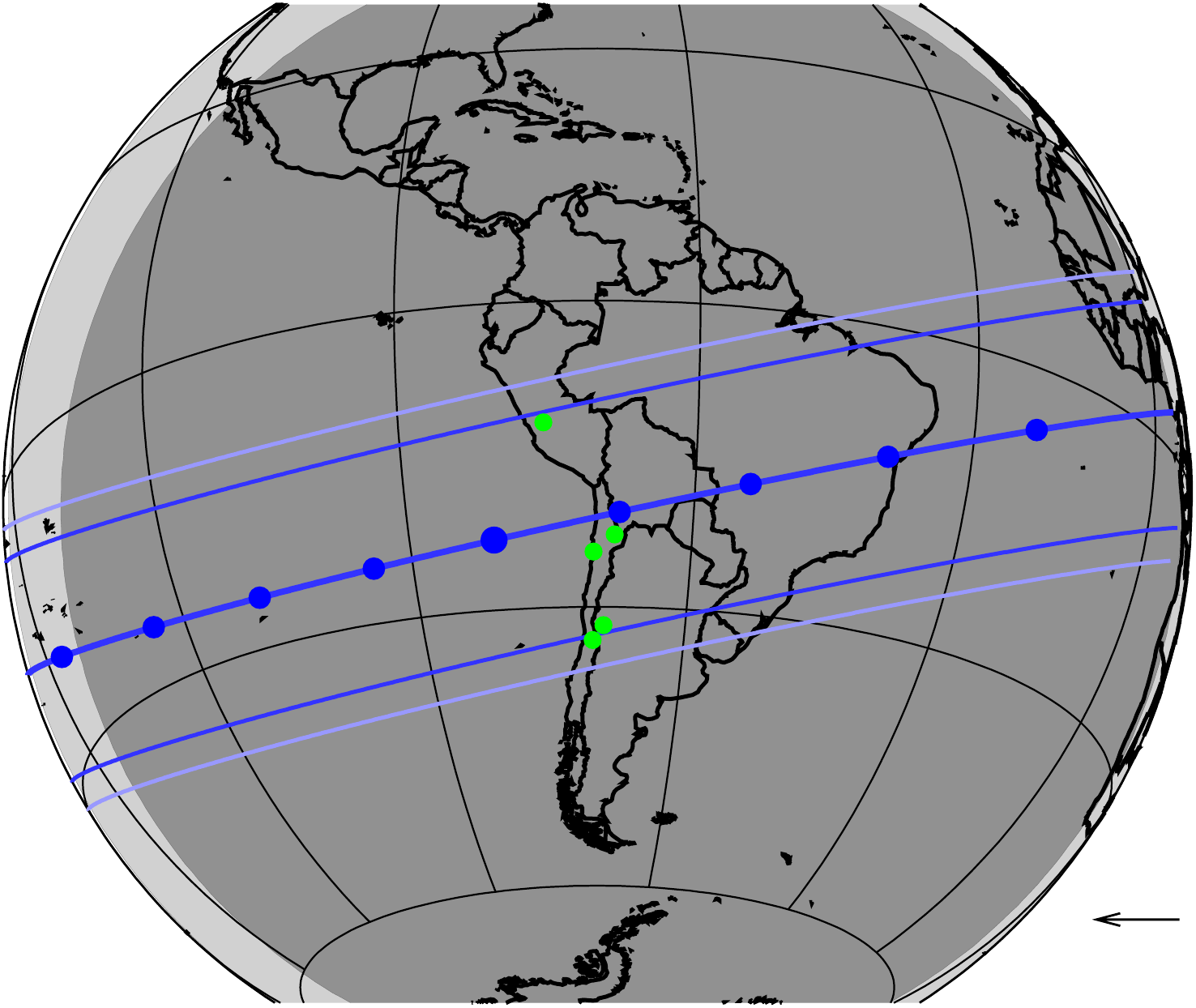}
\subcaption{2012-07-18}
\end{subfigure}%

\begin{subfigure}[b]{.5\linewidth}
\centering \includegraphics[width=0.9\columnwidth]{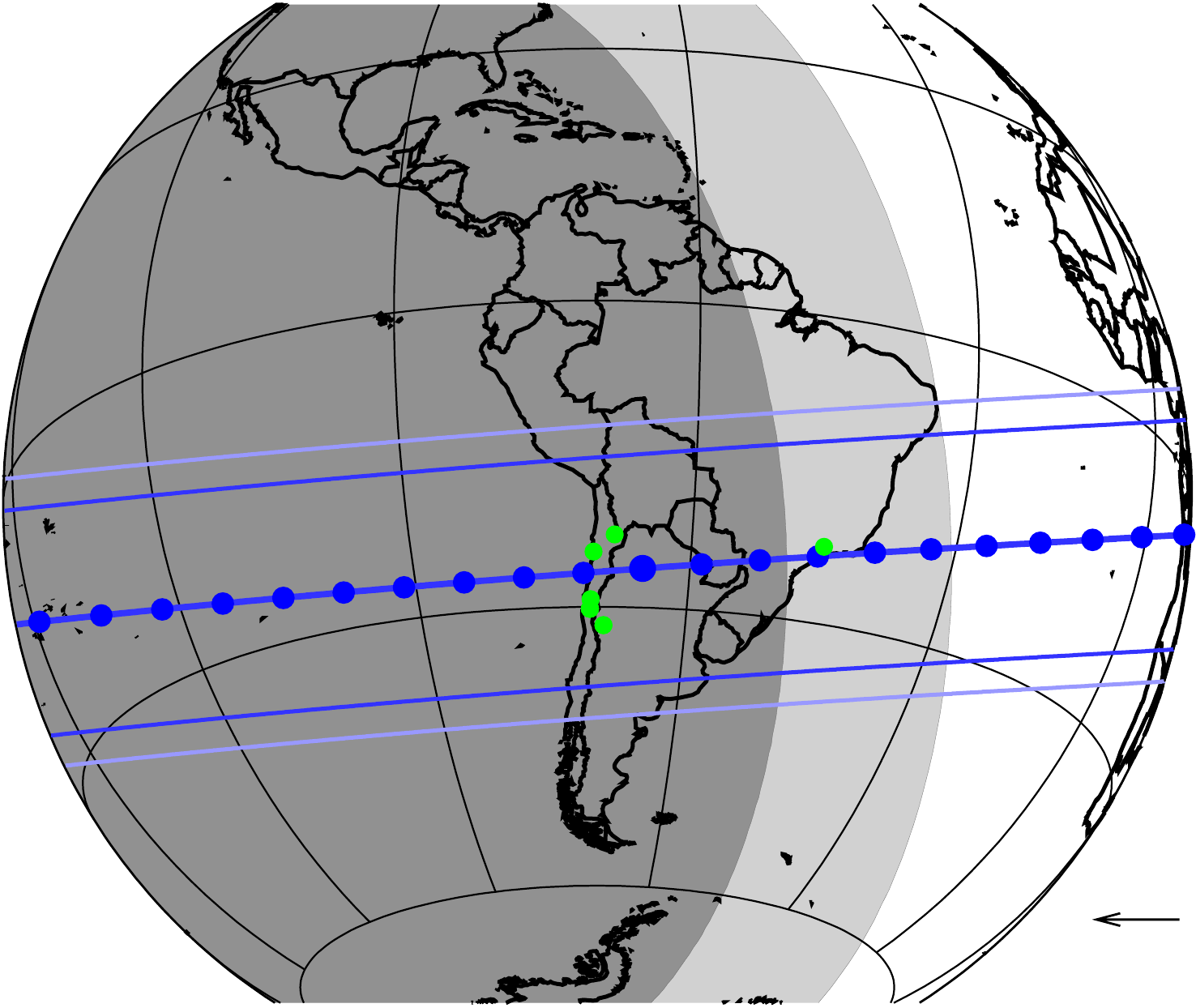}
\subcaption{2013-05-04}
\end{subfigure}%
\begin{subfigure}[b]{.5\linewidth}
\centering \includegraphics[width=0.9\columnwidth]{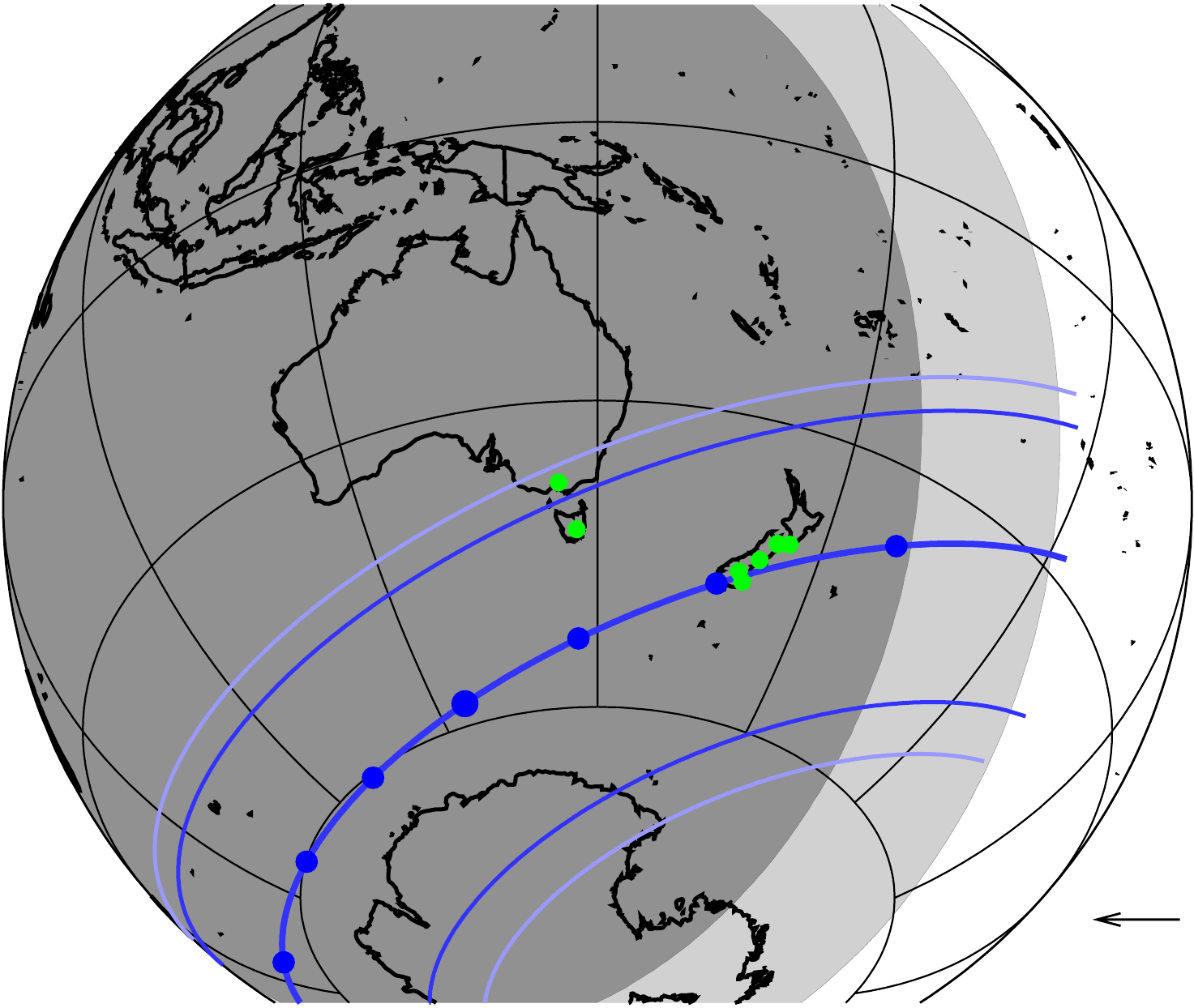}
\subcaption{2015-06-29}
\end{subfigure}%

\begin{subfigure}[b]{.5\linewidth}
\centering \includegraphics[width=0.9\columnwidth]{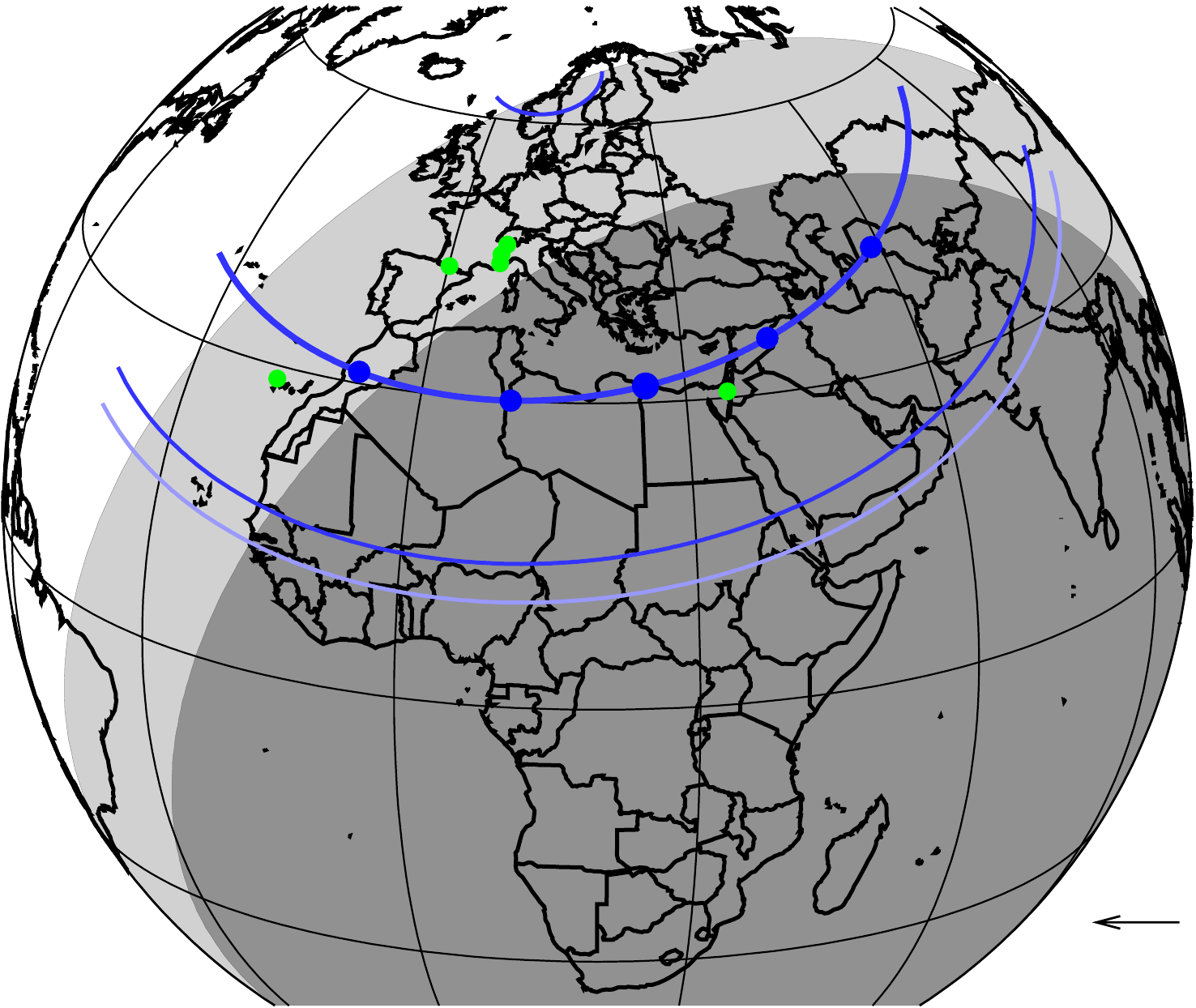}
\subcaption{2016-07-19}
\end{subfigure}%

\caption{%
Reconstruction of Pluto's shadow trajectories on Earth for occultations observed from 2002 to 2016; 
see details in \cite{mez18}. The bullets on the shadow central line are plotted every minute,  and the  black arrow represents the shadow motion direction (see arrow al lower right corner). The dark and light blue thinner lines are the shadow limits corresponding the stellar half-light level and 1~\% stellar drop level (the practical detection limit), respectively. 
The green bullets correspond to the sites with positive detection used in the fit. Areas in dark gray corresponds to full night (Sun elevation below -18$^\circ$) and areas in light grey corresponds to astronomical twilight (Sun elevation between -18$^\circ$ to 0$^\circ$), while day time regions are in white.}%
\label{F:paths}
\end{figure*}



\begin{appendix}\label{S:appendix}
\section{Method to derive astrometric positions from occultation's circumstances}\label{A:method}

We present in this section a method to derive an astrometric position from an occultation's observation, knowing the occultation's circumstances. The determination of an occultation's circumstances consists in computing the Besselian elements. 
The Bessel method makes use of the fundamental plane that passes through the centre of the Earth and perpendicular to the line joining the star and the centre of the object (i.e. the axis of the shadow). 
The method is for example described in \cite{urb13}. The Besselian elements are usually given for the time of conjunction of the star and the object in right ascension but in this paper the reference time is the time of closest angular approach between the star and the object. 

The Besselian elements are $T_0$ the UTC time of the closest approach, $H$ the Greenwich Hour Angle of the star at $T_0$, $x_0$ and $y_0$ the coordinates of the shadow axis at $T_0$ in the fundamental plane, $x'$ and $y'$ the rates of changes in $x$ and $y$ at $T_0$, and $\alpha_s, \delta_s$ the right ascension and the declination of the star. Their computation are fully described in \cite{urb13}.

The quantities $x_0, y_0, x'$ and $y'$ depend on the ephemeris of the body and allow to represent the linear motion of the shadow at the time of the occultation. In this paper, $x_0, y_0$ are expressed in Earth radius unit and $x', y'$ are in Earth radius per day.

From $T_0, \alpha_s, \delta_s$ and $H$, the coordinates\footnotemark[1] of the shadow centre $(\lambda_c,\phi_c)$ at $T_0$ can be derived. 

For an observing site, the method requires the local circumstances which are the mid-time of the occultation and the impact parameter $\rho$, the distance of closest approach between the site and the centre of the shadow in the fundamental plane. Usually, the impact parameter is given in kilometres and when the occultation has only one chord, two solutions (North and South) can be associated. 

The first step is to add a shift to $x_0$ and $y_0$ to take into account the impact parameter, i.e. the fact that the observing site is not right on the centrality of the occultation.

\begin{align}
x_0 & \to  x_0 \pm s \frac{x_0}{\sqrt{x_0^2+y_0^2}} \\
y_0 & \to  y_0 \pm s \frac{y_0}{\sqrt{x_0^2+y_0^2}} 
\end{align}

\noindent where $s$ is the ratio between $\rho$ and Earth radius. 

Given the longitude $\lambda$ and the latitude\footnotemark[1] $\phi$ of the observing site, the coordinates in the fundamental plane are given by :

\footnotetext[1]{Latitude refers to geocentric latitude. Usually coordinates provide geodetic latitude that need to be converted to geocentric latitude.}

\begin{align}
u & =  \cos \phi \sin (\lambda-\lambda_c) \\
v & =  \sin \phi \cos \phi_c - \cos \phi \sin \phi_c \cos (\lambda-\lambda_c) \\
w & =  \sin \phi \sin \phi_c + \cos \phi \cos \phi_c \cos (\lambda-\lambda_c) 
\end{align}

The time of the closest approach for the observer is given by the relation :

\begin{align}
t_{m} & =T_0 + \frac{(u-x_0)x'+(v-y_0) y'}{x'^2+y'^2}
\end{align}

In fact, $t_m, u, v, w$ are calculated iteratively by replacing $\lambda_c$ by $\lambda_c-\Omega (t_m-T_0)$, where $\Omega$ is the rate of Earth's rotation, to take into account the Earth's rotation during $t_m-T_0$.

If $\Delta t$ is the difference between the observed time of the occultation for the observer and the nominal time of the occultation $T_0$, the correction to apply to the Besselian elements $x_0,y_0$ are :

\begin{align}
\Delta x & = (u-x_0)-x' \Delta t \\
\Delta y & = (v-y_0)-y' \Delta t 
\end{align}

The quantities $\Delta x, \Delta y$ are determined iteratively and finally transformed into an offset in right ascension and in declination between the observed occultation and the predicted occultation (from the ephemeris). 

For single chord occultation, there are two solutions (North and South), meaning that we do not 
know whether Pluto’s center went North or South of the star as seen from the observing site. Conversely, for multi-chord occultation there is a unique solution. In that case, the astrometric position deduced from the occultation is the reference ephemeris plus the average offset deduced from all the observing sites. 

This is a powerful method to derive astrometric positions from occultations. It only requires local circumstances of the occultation for the observing sites such as the mid-time of the occultation and the impact parameter. If the impact parameter is not provided, one can deduce it from the timing of immersion and emmersion knowing the size of the object and assuming it is spherical. Thus, the method can be used for any object.

\section{Astrometric positions from other occultations }\label{A:occ}
In this section, we derive astrometric positions from occultations published in various articles using the method previously presented. The Besselian elements corresponding to the occultations are presented in Table~\ref{T:besselian_elements} and the reconstructed shadow trajectories of occultation are presented in Fig.~\ref{F:mapsother}. 

\subsection{Occultation of 9 June 1988}
\cite{mil93} presented the June 9, 1988 Pluto occultation. They derived an astrometric solution by giving the impact parameter for the eight stations that recorded the event.

According to the mid-time of the occultation derived from the paper, we determine the following offsets:

\begin{table}[htp]
\begin{tabular}{crrrrr}
observatory & mid-time & $\rho$ & $\Delta t$ &  $\Delta \alpha \cos \delta$ & $\Delta \delta$ \\
 & (UTC) & (km) & (s) & (mas) & (mas) \\
\hline

Charters Towers & 10:41:27.1$\pm1.23$ &   985 & 130.0 &  20.6 & -33.5 \\
Toowoomba 		& 10:40:50.5$\pm0.55$ &   188 &  93.4 &  18.4 & -33.6 \\
Mt Tamborine 	& 10:40:17.4$\pm0.95$ &   168 &  60.3 &  -4.3 & -33.9 \\
Auckland 		& 10:39:03.3\tablefootmark{1}          &  -687 & -13.8 &  26.6 & -33.9 \\
Hobart 			& 10:41:00.6$\pm1.95$ & -1153 & 103.5 &  19.5 & -33.8 \\
KAO				 & 10:37:26.9$\pm0.15$ &   868 &-110.2 &  19.5 & -33.0 \\
Mt John 		& 10:39:19.6$\pm0.78$ & -1281 &   2.5 &  19.9 & -33.6 \\
\hline
\end{tabular}
\tablefoottext{1}{Uncertainty of timing in Auckland is not provided in \cite{mil93}.}
\end{table}

For Black Birch, there is only the immersion timing so the mid-time of the occultation cannot be derived. The average offset of this occultation was determined using the same set of the preferred astrometric solution of \cite{mil93}, i.e. data from Charters Towers, Hobart, Kuiper Airbone Observatory (KAO) and Mount John.  

Finally, we derive the average offset of $\Delta \alpha \cos \delta=+19.9\pm0.5$ mas and $\Delta \delta=-33.5\pm0.3$ mas.

\subsection{Occultation of 20 July 2002}
\cite{sic03} obtained a light curve of the occultation by Pluto near Arica, North of Chile. They derived an astrometric solution of the occultation by giving distance of closest approach to the centre of Pluto's shadow for Arica  ($975\pm250$km).

In Arica, the mid-time of the occultation occurs at 01:44:03 (UTC), giving $\Delta t=23.2$s. There are two possible solutions but the occultation was also observed at Mami\~na\footnote{The Mami\~na coordinates are $20^\circ04'51.00"$S and $69^\circ12'00.00"$W.} in Chile (Buie, personal communication) so the only possible solution is the South one.

Finally, we derive the offset of $\Delta \alpha \cos \delta=+7.7\pm1.9$ mas and $\Delta \delta=-4.4\pm11.2$ mas, assuming a precision of 2~s for the mid-time.

\subsection{Occultation of 21 August 2002}
\cite{ell03} derived an astrometric solution of the occultation by giving distance of closest approach to the centre of Pluto's shadow for Mauna Kea Observatory ($597\pm32$km) and Lick Observatory ($600\pm32$km). They observed a positive occultation with three telescopes (two in Hawaii and one at Lick Observatory).

As there are at least two stations observing this occultation, there is a unique solution. According to the mid-time of the occultation in the two stations, we derived the following offsets:

\begin{center}
\begin{tabular}{crrrrr}
observatory & mid-time & $\rho$ & $\Delta t$ &  $\Delta \alpha \cos \delta$ & $\Delta \delta$ \\
 & (UTC)  & (km) & (s) & (mas) & (mas) \\
\hline
CFHT 2.2m  & 6:50:33.9$\pm0.5$ & 597 & -598.1 & 16.0 & -8.0 \\
CFHT 0.6m  & 6:50:33.9$\pm1.8$ & 597 & -598.1 & 16.0 & -8.2 \\
Lick obs. & 6:45:48.0$\pm2.8$ & 600 & -884.0 &  14.2  & -11.0 \\ 
\hline
\end{tabular}
\end{center}

Finally, for this occultation, we used an average offset of $\Delta \alpha \cos \delta=+15.4\pm1.0$ mas and $\Delta \delta=-9.1\pm1.7$ mas.

\subsection{Occultation of 12 June 2006}
\cite{you08} presented the analysis of an occultation by Pluto on 12 June 2006. They published the half light time (ingress and egress) and the impact parameter (closest distance to the centre of the shadow) for five stations:
\begin{itemize}
\item REE = Reedy Creek Observatory, QLD, AUS (0.5 m aperture).
\item AAT = Anglo-Australian Observatory, NSW, AUS (4 m).
\item STO = Stockport Observatory, SA, AUS (0.5 m).
\item HHT = Hawkesbury Heights, NSW, AUS (0.2 m).
\item CAR = Carter Observatory, Wellington, NZ (0.6 m)
\end{itemize}

These parameters allow us to compute the mid-time of the occultation and to finally derive an offset for each station:

\begin{center}
\begin{tabular}{crrrrr}
observatory & mid-time & $\rho$ & $\Delta t$ &  $\Delta \alpha \cos \delta$ & $\Delta \delta$ \\
 & (UTC)  & (km) & (s) & (mas) & (mas) \\
\hline
REE & 16:23:00.64$\pm2.61$ &  836.6 & -125.2 &  9.4 & -0.5 \\
AAT & 16:23:19.67$\pm0.05$ &  571.8 & -106.1 &  9.6 & -0.5 \\
STO & 16:23:59.62$\pm0.80$ &  382.2 &  -66.2 &  9.7 & -0.5 \\
HHT & 16:23:17.70$\pm2.12$ &  302.5 & -108.1 &  9.1 & -0.4 \\
CAR & 16:22:30.82$\pm1.96$ & -857.6 & -155.0 & 11.2 & -0.4 \\
\hline
\end{tabular}
\end{center}

Finally, for this occultation, we used an average offset of $\Delta \alpha \cos \delta=+9.8\pm0.8$ mas and $\Delta \delta=-0.4\pm0.1$ mas.

\subsection{Occultation of 18 March 2007}
\cite{per08} presented an analysis of an occultation by Pluto observed in several places in USA on 18 March 2007. 

From five stations, they derived the geometry of the event by providing the mid-time (UTC) of the event at 10:53:49$\pm$00:01 (giving $\Delta t=-344.1$s) and an impact parameter of $1319\pm4$ km for the Multiple Mirror Telescope Observatory (MMTO).

According to the geometry of the event, the South solution ($\rho=-1319$ km) has to be adopted, giving the offset related to JPL DE436/PLU055 ephemeris of $\Delta \alpha \cos \delta =  10.7\pm0.3$ mas and $\Delta \delta = 0.8\pm0.2$mas.

\subsection{Occultation of 23 June 2011}
\cite{gul15} presented a grazing occultation by Pluto observed in IRTF (Mauna Kea Observatory) on 23 June 2011. They derived an impact parameter of $1138\pm3$ km and a mid-time (UTC) of the event at 11:23:03.07 ($\pm0.10$~s).

The single chord leads to two possible solutions providing the following offset related to JPL DE436/PLU055 ephemeris: \\
\begin{center}
\begin{tabular}{crr}
  & North  & South  \\ 
\hline
$\Delta \alpha \cos \delta$ (mas) &  16.1 & 5.3 \\ 
$\Delta \delta$ (mas) &  5.5 & 106.1\\
\hline
\end{tabular}
\end{center}

According to \cite{gul15}, the North solution has to be adopted. Finally, the offset is $\Delta \alpha \cos \delta =  16.1\pm0.1$ mas and $\Delta \delta = 5.5\pm0.1$mas, assuming the estimated precision of the timing and the impact parameter.

\subsection{Occultation of 04 May 2013}
\cite{olk15} presented the occultation by Pluto on 4 May 2013 observed in South America.

They derived the mid-time (UTC) of the event at 08:23:21.60$\pm0.05$s (giving $\Delta t=99.8$s) and an impact parameter of $370\pm5$ km for the LCOGT at Cerro Tololo.

From these circumstances, we derived an offset related to JPL DE436/PLU055 ephemeris of $\Delta \alpha \cos \delta =  18.7\pm0.1$ mas and $\Delta \delta = 8.4\pm0.2$mas

\subsection{Occultation of 23 July 2014}
\cite{pas16} published the observation of two single-chord occultations at Mont John (New Zealand) on June 2014. They provided the timing and impact parameter for the two occultations. 

The fitted impact parameter for 23 July is $\rho=480\pm120$km providing two possible solutions and the mid-time (UTC) of the occultation 14:24:31$\pm4$s is derived from the ingress and egress times at 50\% and corresponds to $\Delta t = -88.1$s.

Each solution provides the following offset related to JPL DE436/PLU055 ephemeris: \\
\begin{center}
\begin{tabular}{crr}
  & North  & South  \\ 
\hline
$\Delta \alpha \cos \delta$ (mas) &  30.3 & 22.9 \\ 
$\Delta \delta$ (mas) &  3.7 & 44.9\\
\hline
\end{tabular}
\end{center}

According to the precisions of the mid-time and of the impact parameter, the estimated precision of the offset is 4.0 mas for $\Delta \alpha \cos \delta$ and 5.2 mas for $\Delta \delta$.

\subsection{Occultation of 24 July 2014}
\cite{pas16} also provided circumstances of the occultation on 24 July 2014 at Mont John Observatory.

The fitted impact parameter is $\rho=510\pm140$km providing two possible solutions and the mid-time (UTC) of the occultation 11:42:29$\pm8$s is derived from the ingress and egress times at 50\% and corresponds to $\Delta t = 9.1$s.

Each solution provides the following offset related to JPL DE436/PLU055 ephemeris: \\
\begin{center}
\begin{tabular}{crr}
  & North  & South  \\ 
\hline
$\Delta \alpha \cos \delta$ (mas) &  3.4 & 11.3 \\ 
$\Delta \delta$ (mas) &  29.1 & -14.6\\
\hline
\end{tabular}
\end{center}

According to the precisions of the mid-time and of the impact parameter, the estimated precision of the offset is 7.7 mas for $\Delta \alpha \cos \delta$ and 6.1 mas for $\Delta \delta$.

\subsection{Occultation of 29 June 2015}
\cite{pas17} presented the occultation by Pluto on 29 June 2015.

They derived the mid-time (UTC) of the event at 16:52:50 (giving $\Delta t=-111.4$s) and an impact parameter of $-53.1$ km for the Mont John Observatory in New Zealand.

From these circumstances, we derived an offset of $\Delta \alpha \cos \delta=22.1$mas and $\Delta \delta =12.7$mas related to JPL DE436/PLU055 ephemeris. The precision of the offset cannot be determined since the precision in mid-time and in the impact parameter are not indicated.

\begin{table*}[htp!]
\begin{center}
\caption{Besselian elements for occultations listed in the appendix derived with Gaia DR2 for the star's position and JPL DE436/PLU055 for Pluto's ephemeris.}
\label{T:besselian_elements}
\small{\begin{tabular}{crrrrrrr}
\hline
$T_0$ & $x_0$ & $y_0$ & $x'$ & $y'$ & $H$ & $\alpha_s$ & $\delta_s$ \\
\hline
 1988-06-09 10:39:17.1 &  0.006535856 &  -0.390599080 & -242.990271254 &   -4.176391160 & -47.003163462  & 223.041508925 &  0.750884462 \\
2002-07-20 01:43:39.8 & -0.015137748 &  0.078729716 & -221.595155776 & -42.613814665 &  45.303191676 & 255.075123563 & -12.694996935 \\
2002-08-21 07:00:32.0 & 0.091629552 &  -0.047418125 & -41.470159949 & -80.186411178 &  -27.314474978 & 254.705972362 & -12.858853587 \\
2006-06-12 16:25:05.8 & 0.008081468 & -0.393907343 & -320.357408358 &   -6.588025106 &  39.386450596 & 265.300310118 & -15.692941450 \\
2007-03-18 10:59:33.1 & -0.283497691 &  0.985999061 & 92.267892934 & 26.509008184 &  -58.153737570 & 268.773723165 & -16.476135950 \\
2011-06-23 11:23:48.2 & -0.043316318 &  0.403059932 & -320.782100593 & -34.487845936 &  50.562763031 & 276.481160400 & -18.801937982 \\
2013-05-04 08:21:41.8 & 0.013860759 &  -0.136954904 &-137.646799082 &  -13.969616086 & 16.003277103 & 281.968884350 & -19.690120815 \\
2014-07-23 14:25:59.1 & 0.110372760 & -0.614706119 & -300.130385882 &   -53.903828467 &  -20.940785660 & 282.382245191 & -20.373331983 \\
2014-07-24 11:42:19.9 & 0.075661748 & -0.419500350 & -297.988040527 &   -53.754831391 &  -22.209299195 & 282.360471376 & -20.376972931 \\
2015-06-29 16:54:41.4 & 0.106938572 &  -0.628240925 & -318.341422110 &  -54.232339089 &  -2.294494383 & 285.206150857 & -20.694717628 \\
\hline
\end{tabular}}
\tablefoot{$T_0$ is the UTC time of the closest approach, $x_0,y_0$ are the coordinates of the shadow axis in the fundamental plane at $T_0$ (in Earth's radius unit), $x',y'$ are the rate of change in $x$ and $y$ at $T_0$ (in equatorial Earth's radius per day), $H$ is the Greenwich Hour Angle of the star at $T_0$ (in degrees), and $\alpha_s,\delta_s$ are the right ascension and declination of the star (in degrees).}
\end{center}
\end{table*}

\begin{figure*}
\begin{subfigure}[b]{.5\linewidth}
\centering \includegraphics[width=0.9\columnwidth]{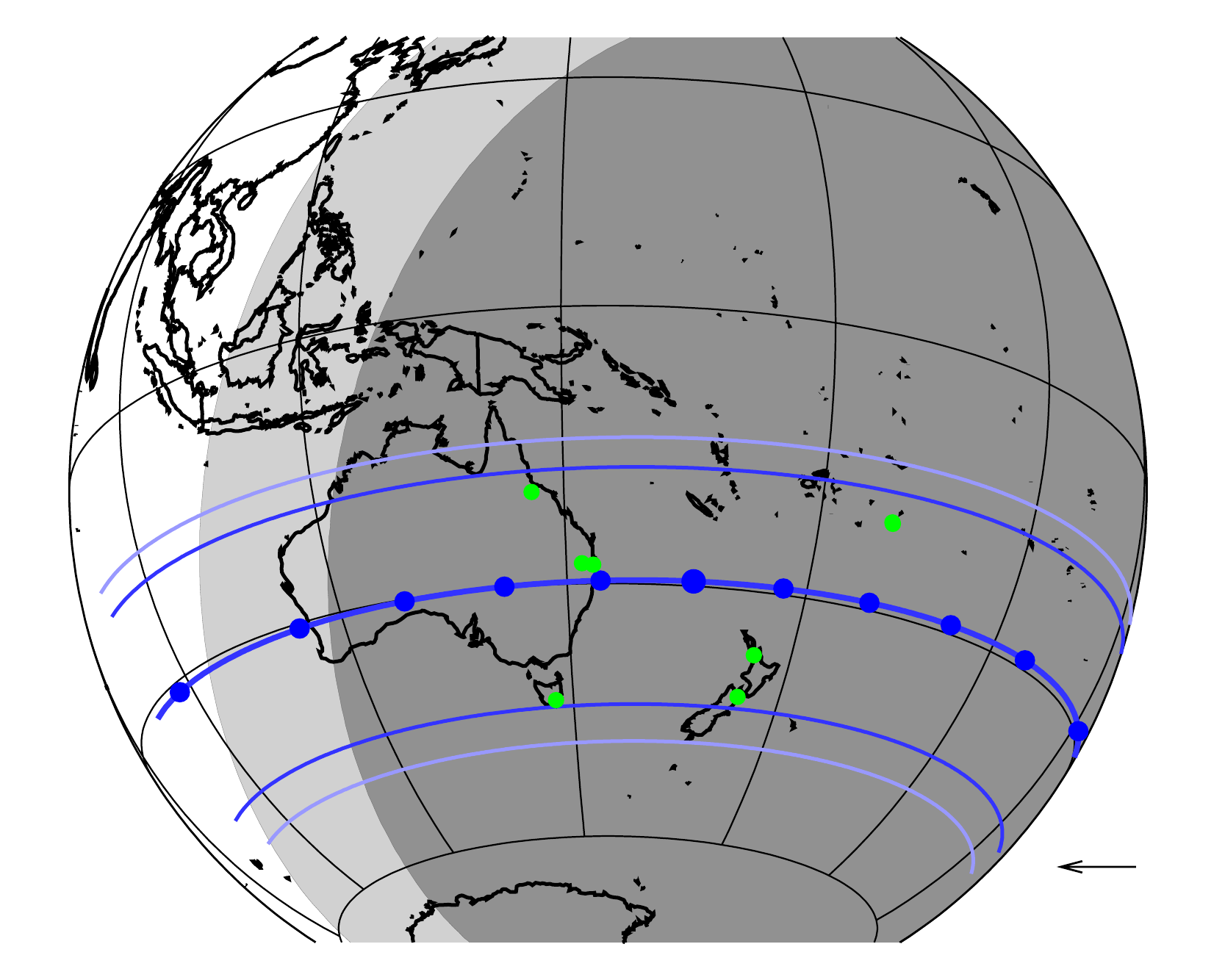}
\subcaption{1988-06-09}
\end{subfigure}%
\begin{subfigure}[b]{.5\linewidth}
\centering \includegraphics[width=0.9\columnwidth]{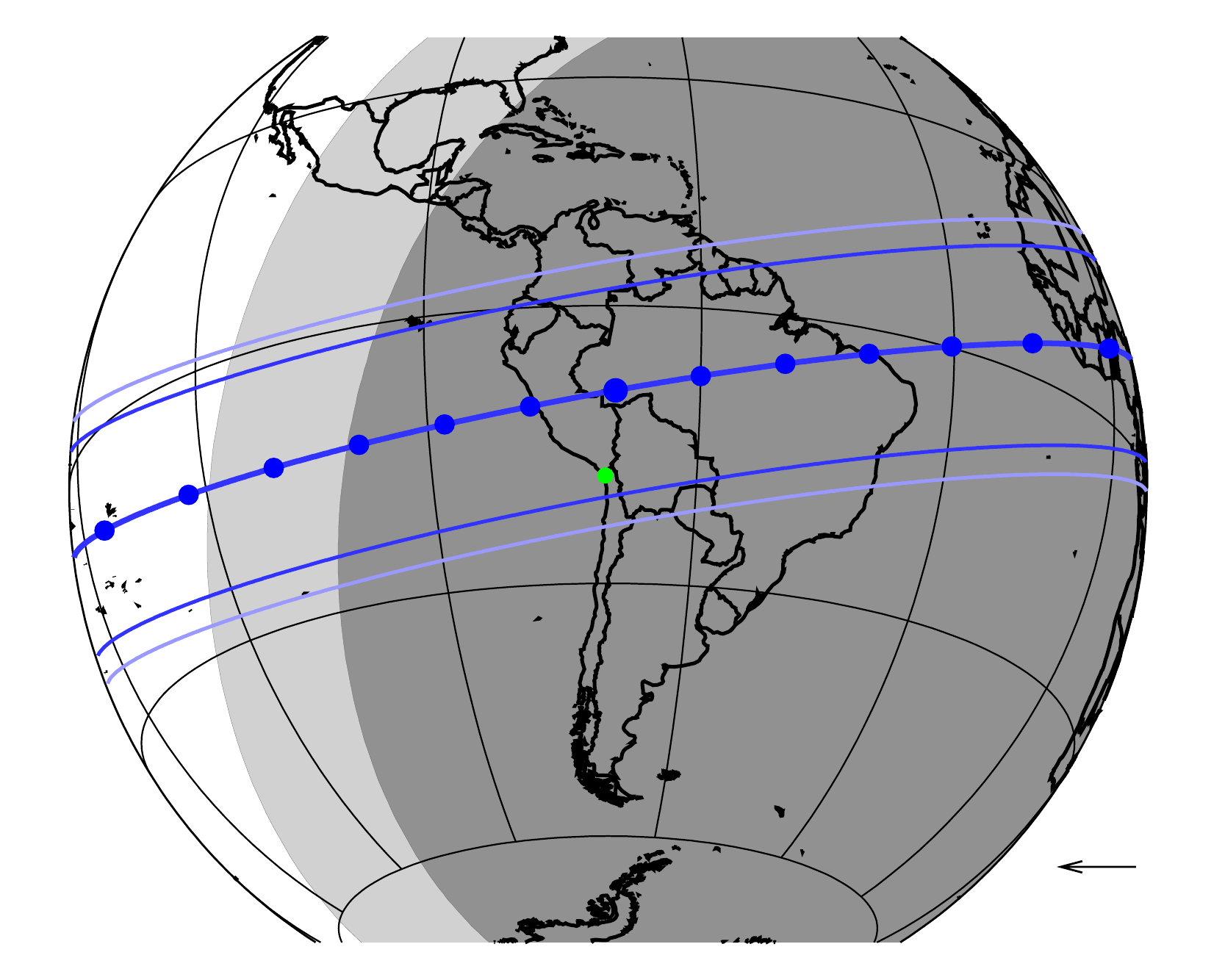}
\subcaption{2002-07-20}
\end{subfigure}%

\begin{subfigure}[b]{.5\linewidth}
\centering \includegraphics[width=0.9\columnwidth]{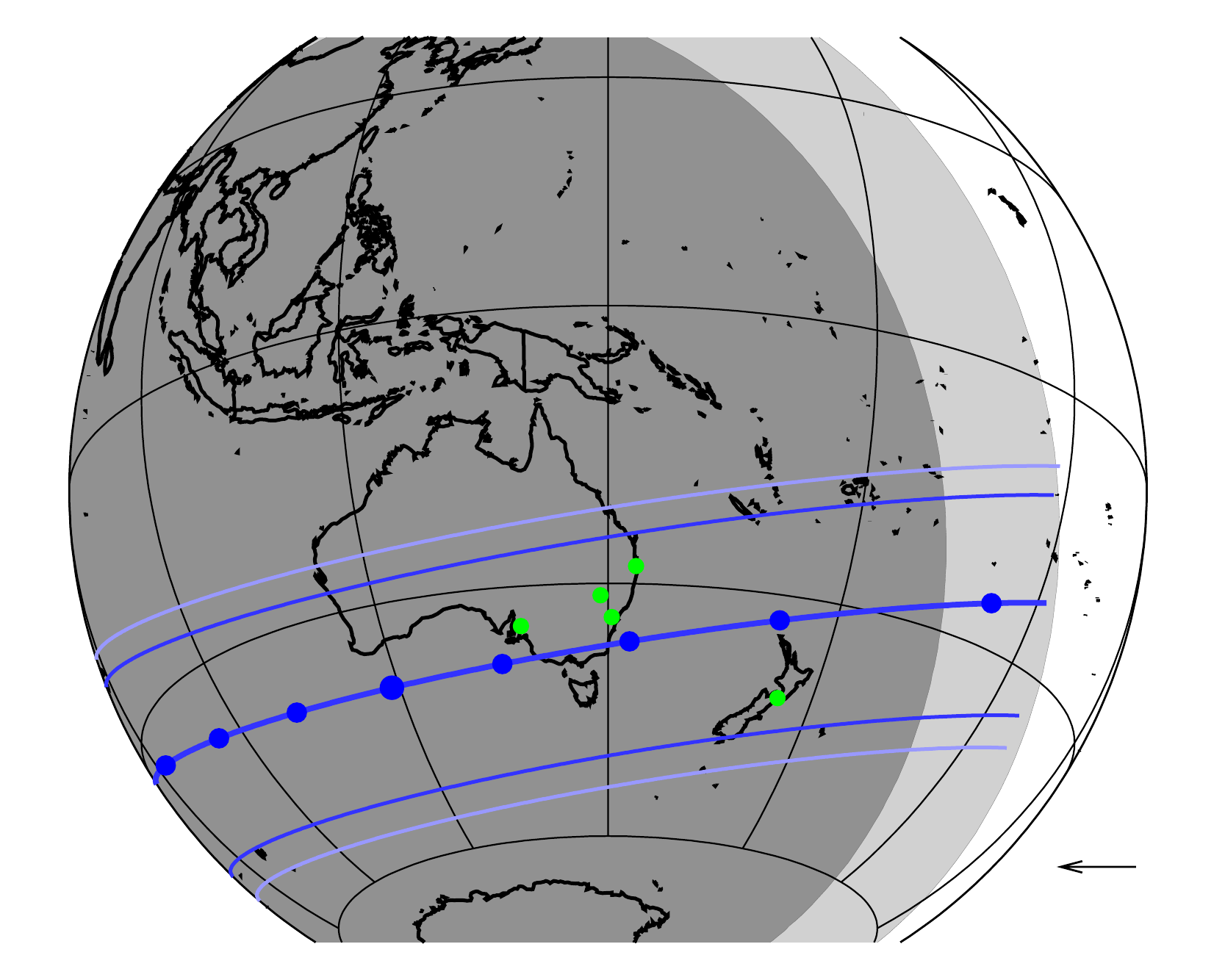}
\subcaption{2006-06-12}
\end{subfigure}%
\begin{subfigure}[b]{.5\linewidth}
\centering \includegraphics[width=0.9\columnwidth]{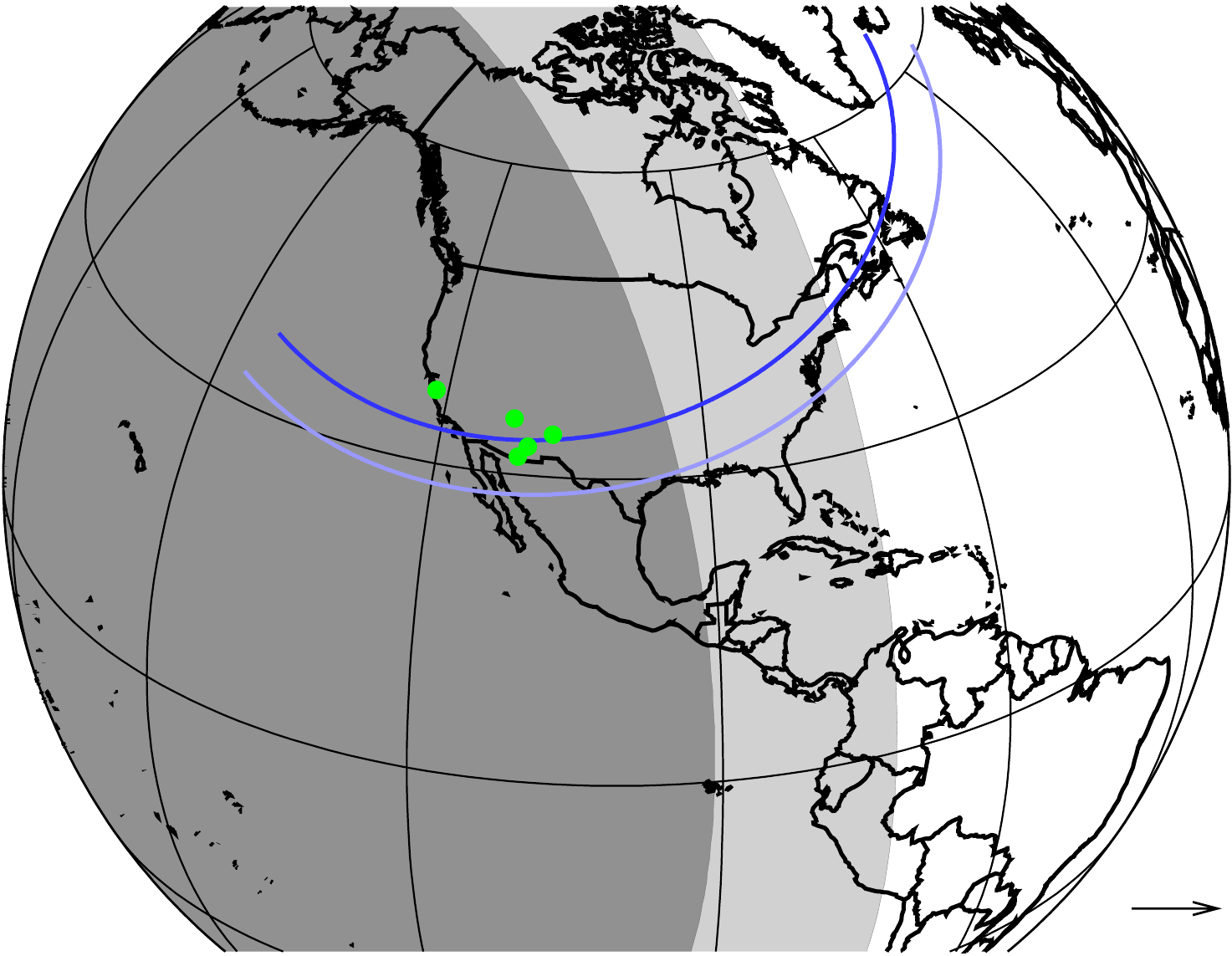}
\subcaption{2007-03-18}
\end{subfigure}%

\begin{subfigure}[b]{.5\linewidth}
\centering \includegraphics[width=0.9\columnwidth]{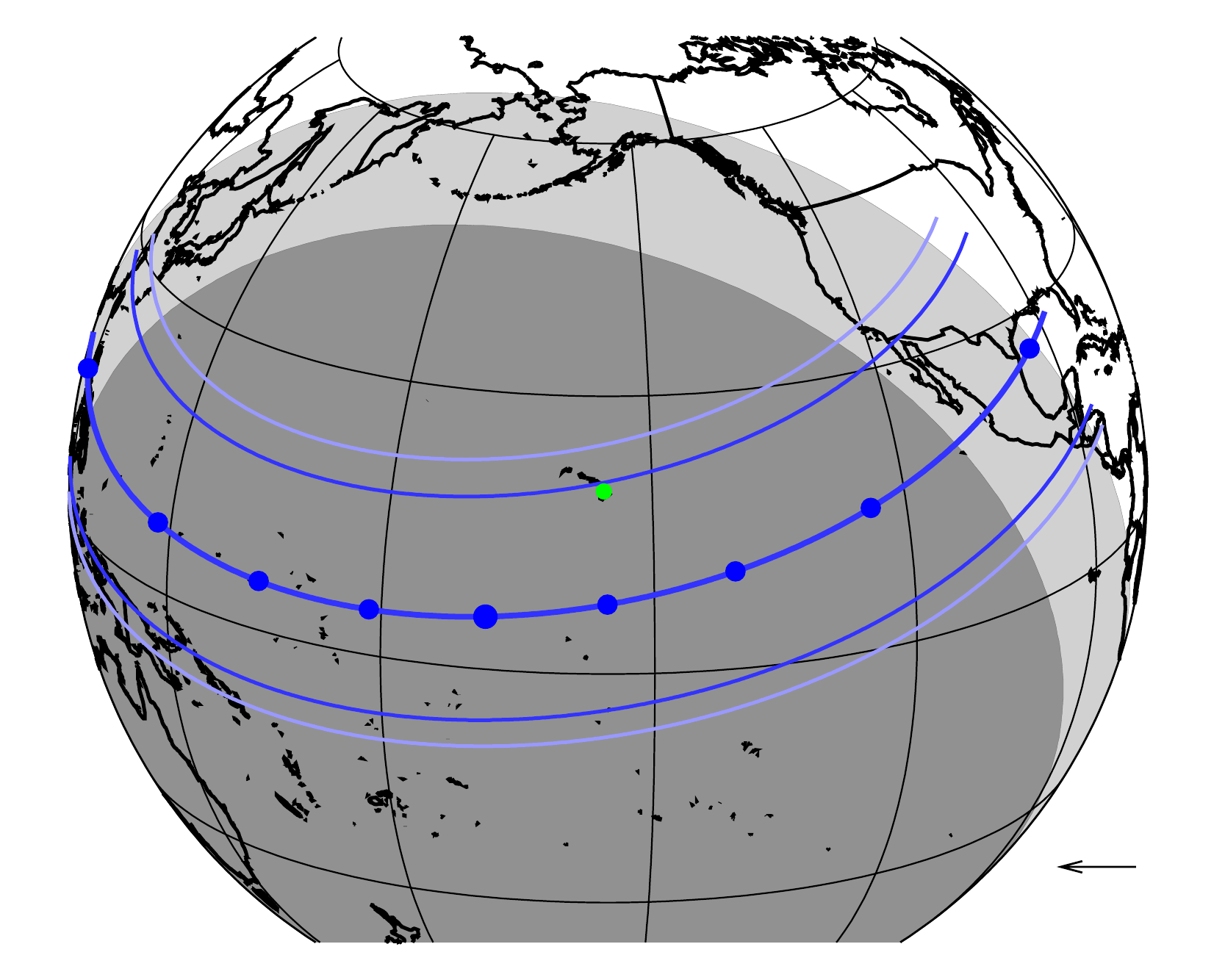}
\subcaption{2011-06-23}
\end{subfigure}%
\begin{subfigure}[b]{.5\linewidth}
\centering \includegraphics[width=0.9\columnwidth]{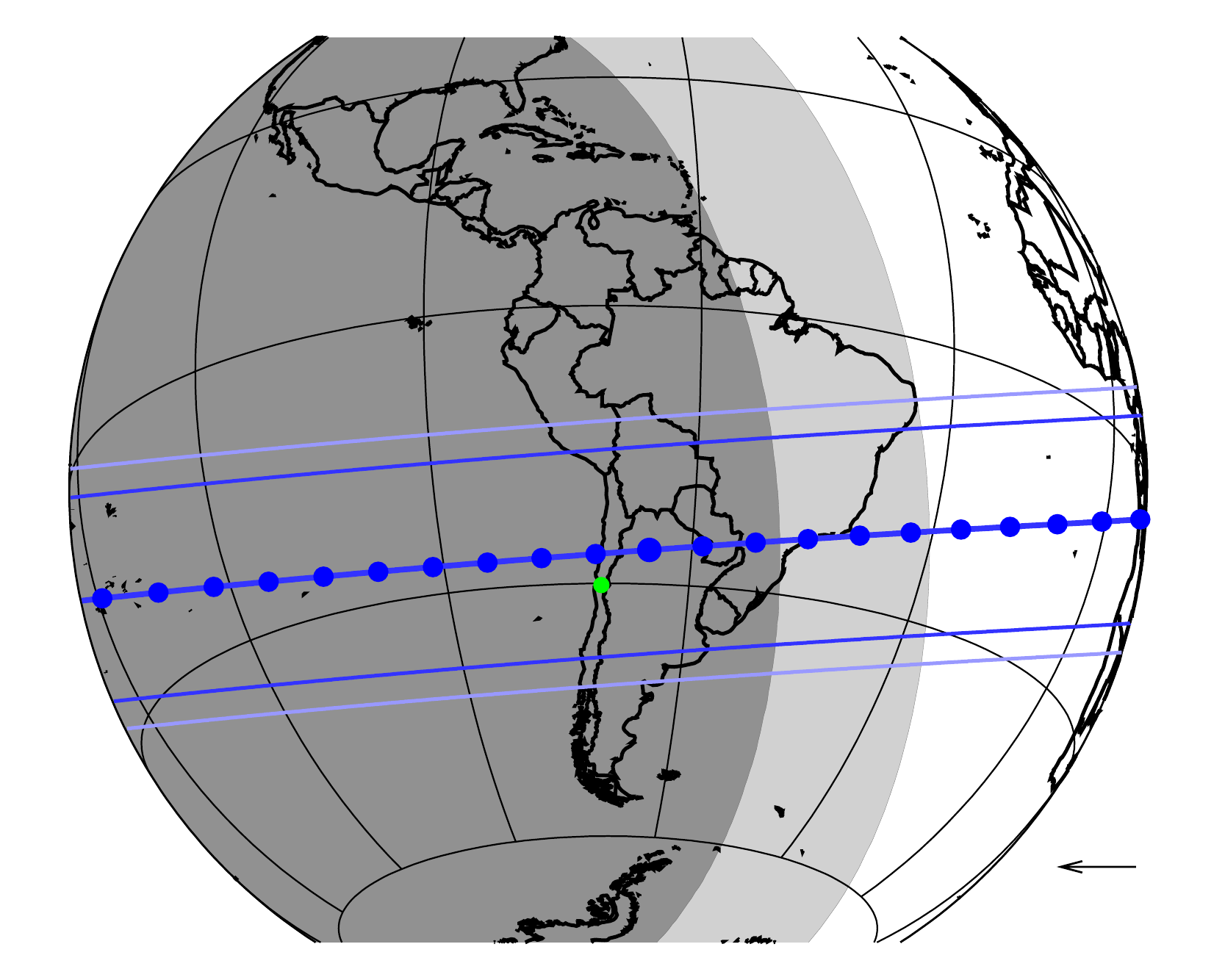}
\subcaption{2013-05-04}
\end{subfigure}%

\end{figure*}
\begin{figure*}[htb]\ContinuedFloat

\begin{subfigure}[b]{.5\linewidth}
\centering \includegraphics[width=0.9\columnwidth]{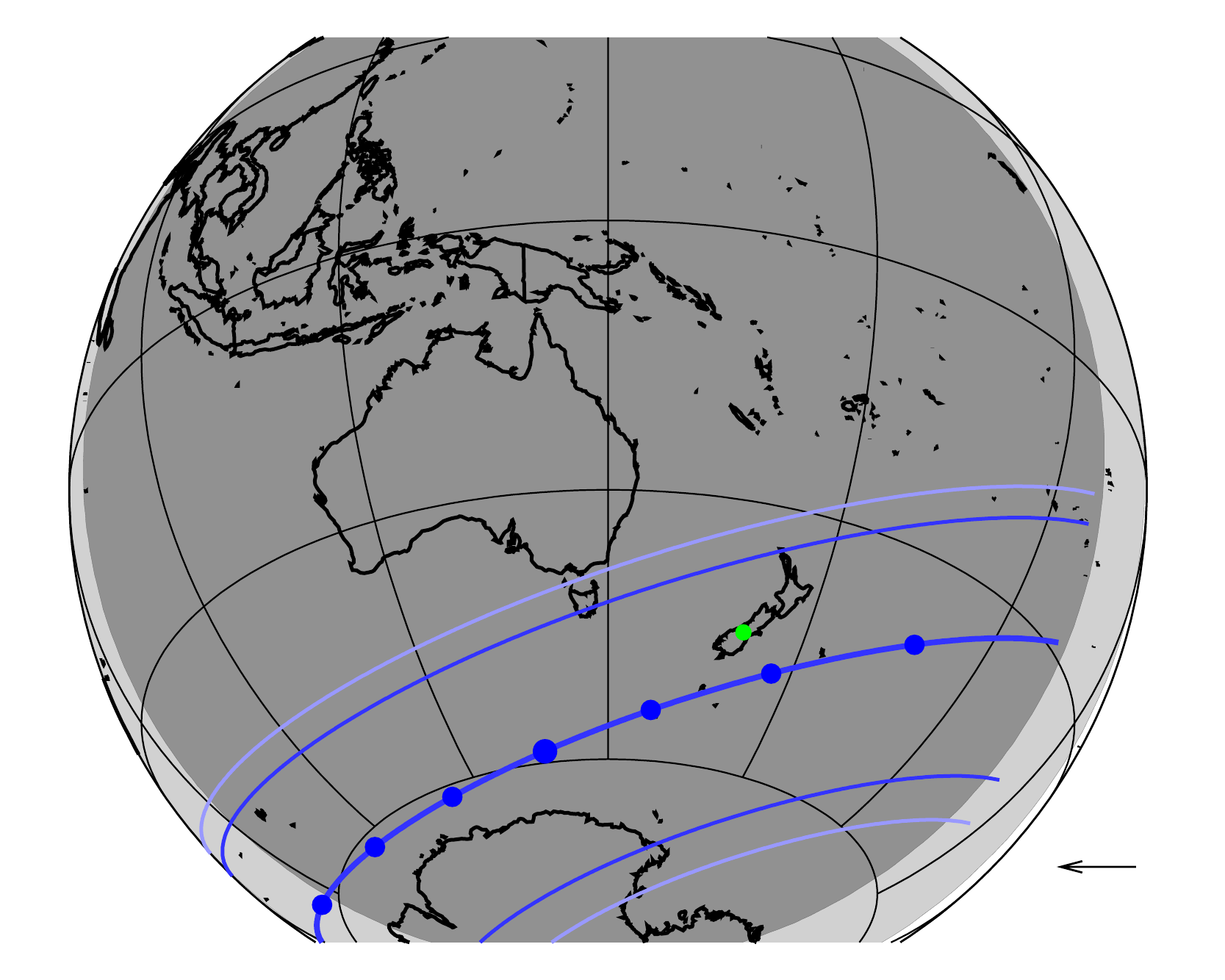}
\subcaption{2014-07-23 (North solution)}
\end{subfigure}%
\begin{subfigure}[b]{.5\linewidth}
\centering \includegraphics[width=0.9\columnwidth]{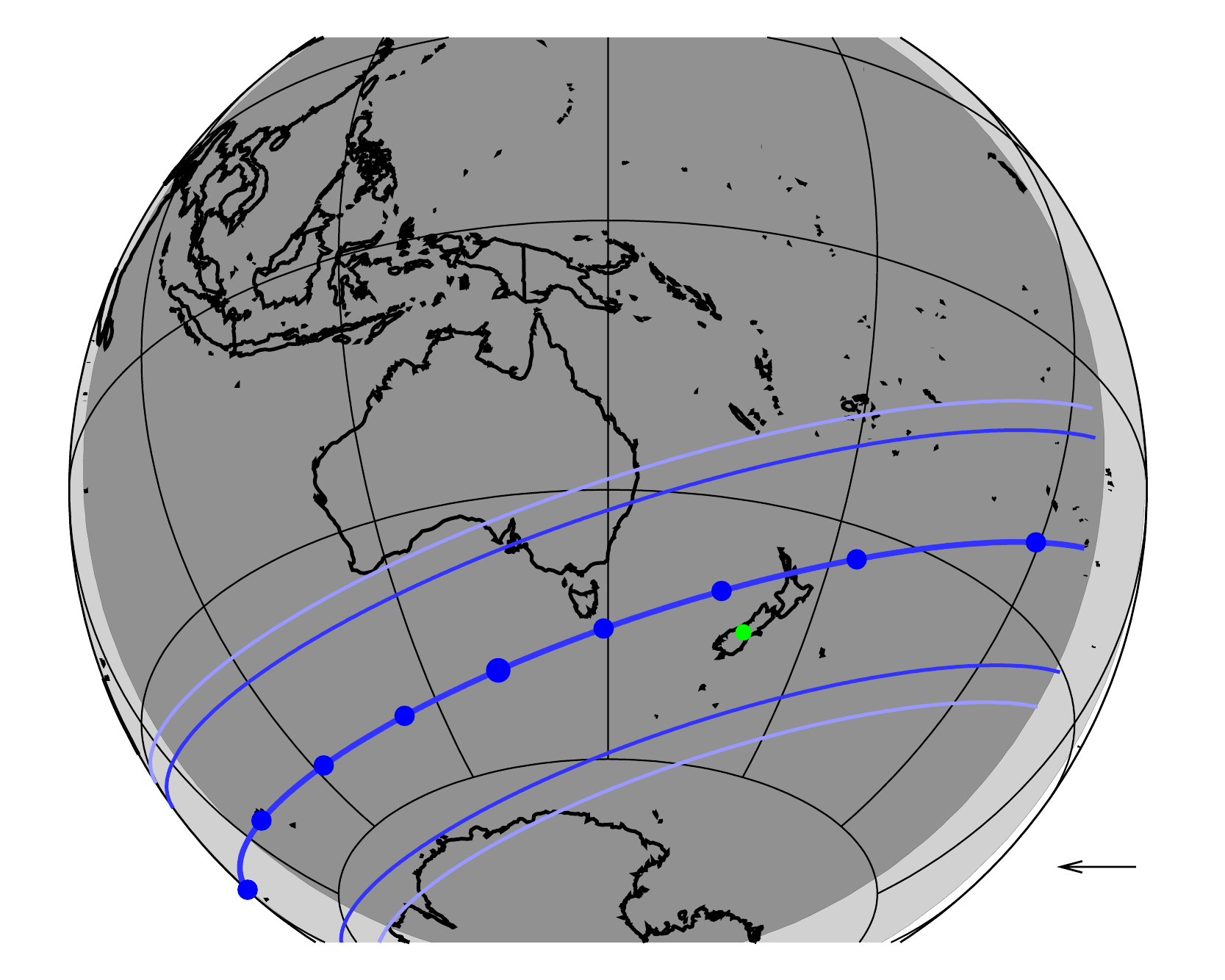}
\subcaption{2014-07-23 (South solution)}
\end{subfigure}%

\begin{subfigure}[b]{.5\linewidth}
\centering \includegraphics[width=0.9\columnwidth]{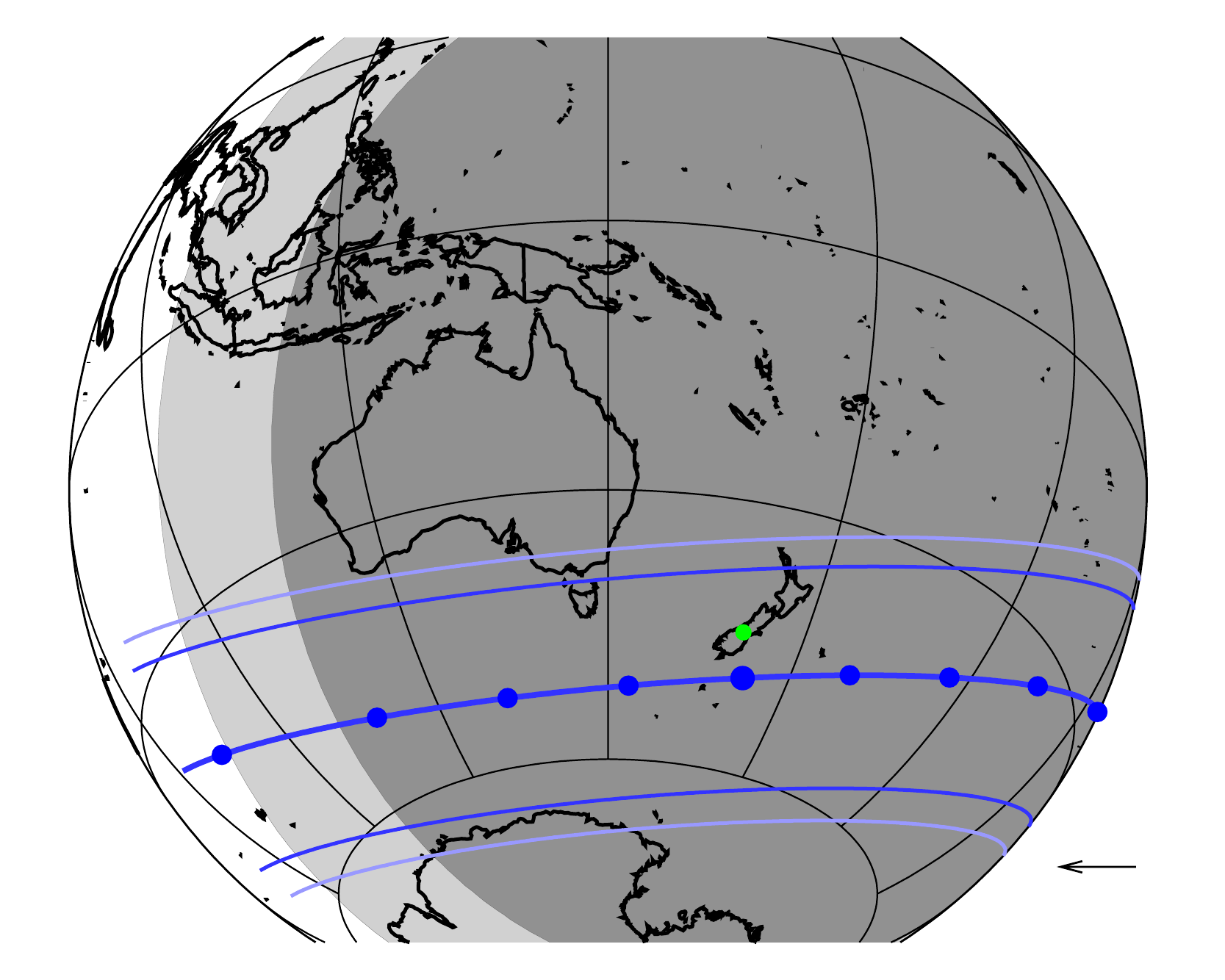}
\subcaption{2014-07-24 (North solution)}
\end{subfigure}%
\begin{subfigure}[b]{.5\linewidth}
\centering \includegraphics[width=0.9\columnwidth]{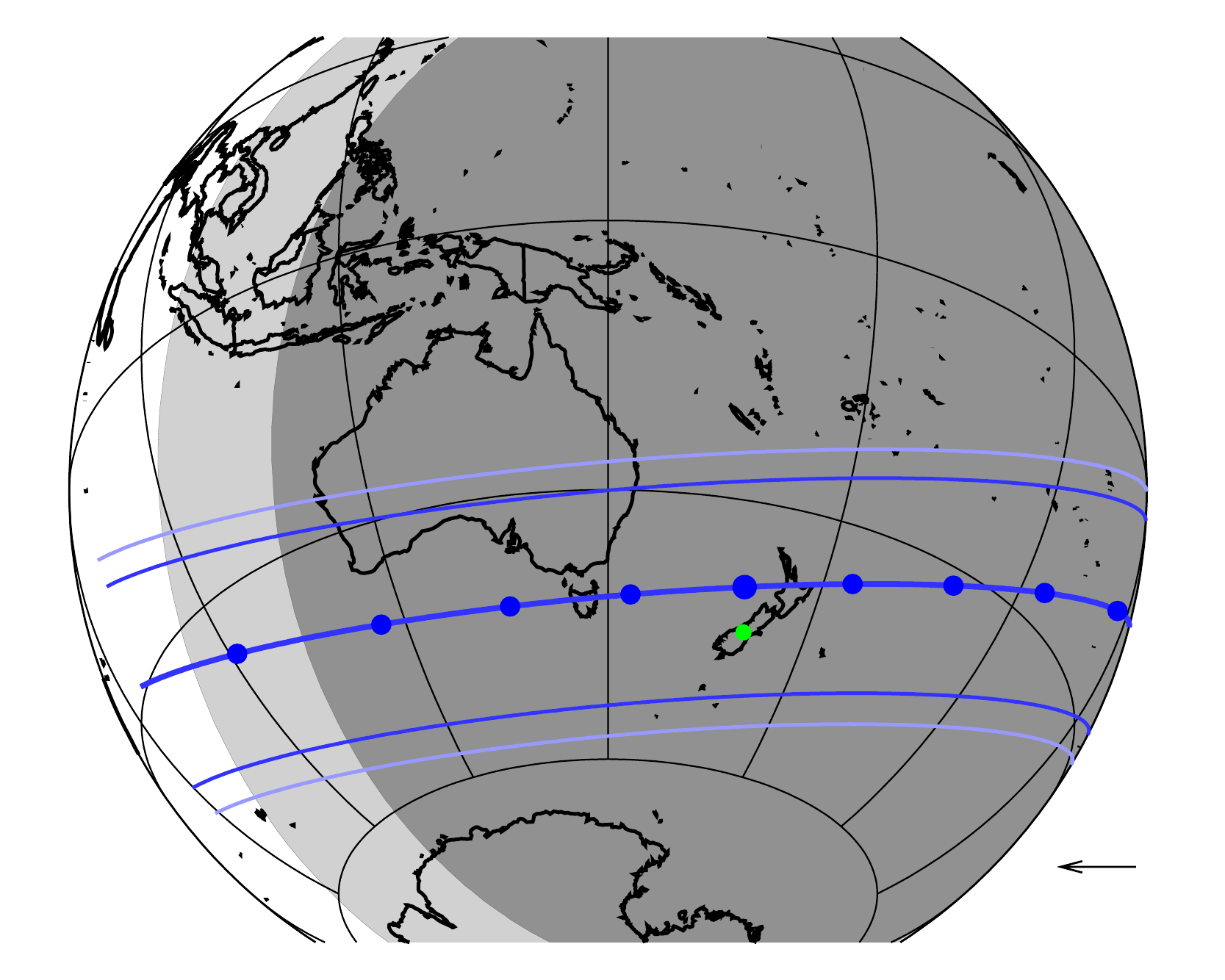}
\subcaption{2014-07-24 (South solution)}
\end{subfigure}%

\begin{subfigure}[b]{.5\linewidth}
\centering \includegraphics[width=0.9\columnwidth]{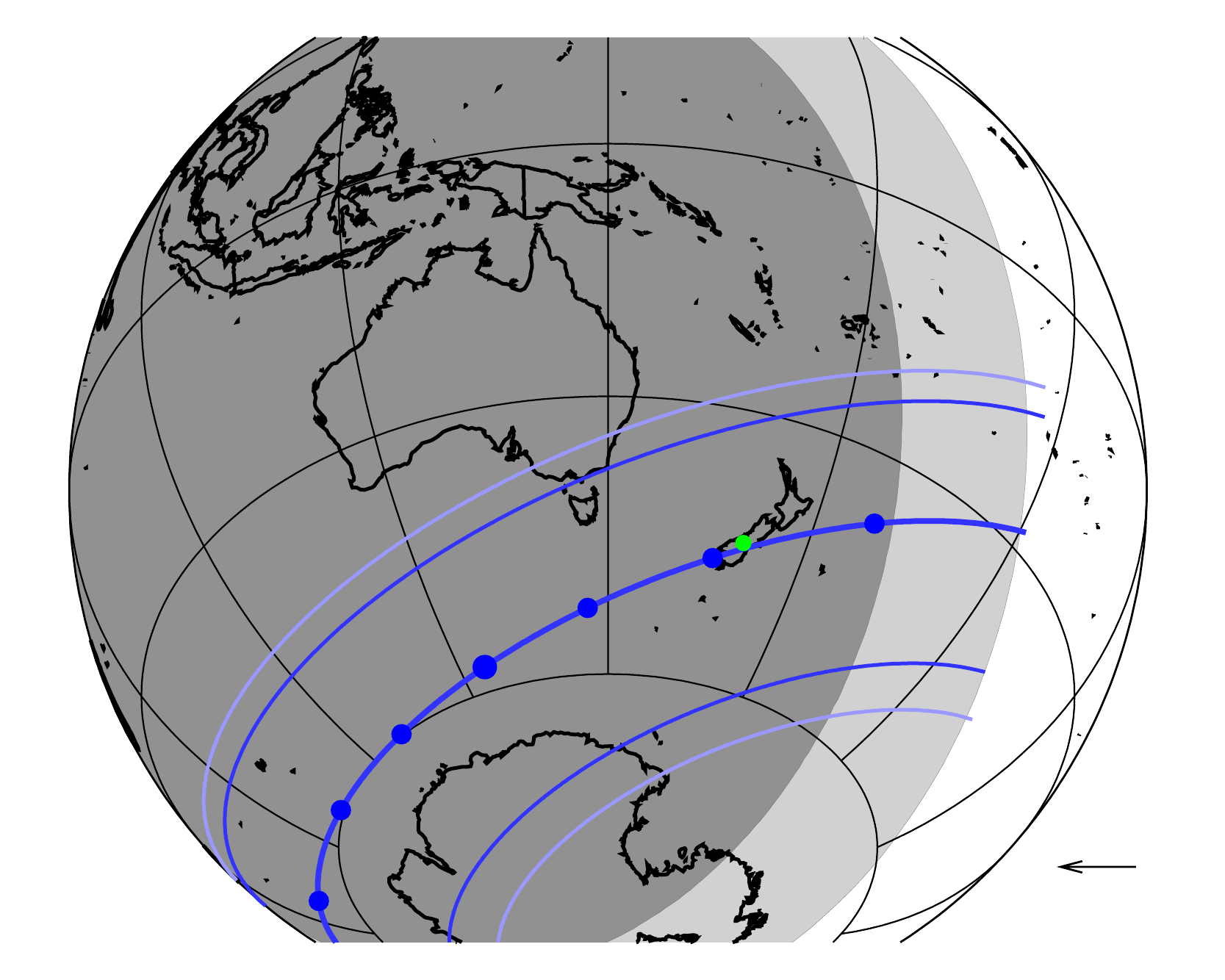}
\subcaption{2015-06-29}
\end{subfigure}%
\caption{%
Reconstruction of Pluto's shadow trajectories on Earth for occultations presented in other publications from 1988 to 2015. The legend is similar to Fig.\ref{F:paths}. 
}%
\label{F:mapsother}
\end{figure*}


\end{appendix} 

\end{document}